

%
%
%
\def\unredoffs{} \def\redoffs{\voffset=-.31truein\hoffset=-.59truein}
\def\speclscape{\special{ps: landscape}}
%
%
%
%
\newbox\leftpage \newdimen\fullhsize \newdimen\hstitle \newdimen\hsbody
\tolerance=1000\hfuzz=2pt
\catcode`\@=11 
\def\bigans{b }
\def\answ{b }

%
\ifx\answ\bigans\message{(This will come out unreduced.}
\magnification=1200\unredoffs\baselineskip=16pt plus 2pt minus 1pt
\hsbody=\hsize \hstitle=\hsize 
\else\message{(This will be reduced.} \let\l@r=L
\magnification=1000\baselineskip=16pt plus 2pt minus 1pt \vsize=7truein
\redoffs \hstitle=8truein\hsbody=4.75truein\fullhsize=10truein\hsize=\hsbody
\output={\ifnum\pageno=0 
  \shipout\vbox{\speclscape{\hsize\fullhsize\makeheadline}
    \hbox to \fullhsize{\hfill\pagebody\hfill}}\advancepageno
  \else
  \almostshipout{\leftline{\vbox{\pagebody\makefootline}}}\advancepageno
  \fi}
\def\almostshipout#1{\if L\l@r \count1=1 \message{[\the\count0.\the\count1]}
      \global\setbox\leftpage=#1 \global\let\l@r=R
 \else \count1=2
  \shipout\vbox{\speclscape{\hsize\fullhsize\makeheadline}
      \hbox to\fullhsize{\box\leftpage\hfil#1}}  \global\let\l@r=L\fi}
\fi
%
\newcount\yearltd\yearltd=\year\advance\yearltd by -1900

\def\Title#1#2{\nopagenumbers\abstractfont\hsize=\hstitle\rightline{#1}%
\vskip 1in\centerline{\titlefont #2}\abstractfont\vskip .5in\pageno=0}
\def\Date#1{\vfill\leftline{#1}\tenpoint\supereject\global\hsize=\hsbody%
\footline={\hss\tenrm\folio\hss}}
%

\def\draftmode{\message{ DRAFTMODE }\def\draftdate{{\rm preliminary draft:
\number\month/\number\day/\number\yearltd\ \ \hourmin}}%
\headline={\hfil\draftdate}\writelabels\baselineskip=20pt plus 2pt minus 2pt
 {\count255=\time\divide\count255 by 60 \xdef\hourmin{\number\count255}
  \multiply\count255 by-60\advance\count255 by\time
  \xdef\hourmin{\hourmin:\ifnum\count255<10 0\fi\the\count255}}}
\def\nolabels{\def\wrlabeL##1{}\def\eqlabeL##1{}\def\reflabeL##1{}}
\def\writelabels{\def\wrlabeL##1{\leavevmode\vadjust{\rlap{\smash%
{\line{{\escapechar=` \hfill\rlap{\sevenrm\hskip.03in\string##1}}}}}}}%
\def\eqlabeL##1{{\escapechar-1\rlap{\sevenrm\hskip.05in\string##1}}}%
\def\reflabeL##1{\noexpand\llap{\noexpand\sevenrm\string\string\string##1}}}
\nolabels
%
\global\newcount\secno \global\secno=0
\global\newcount\meqno \global\meqno=1
\def\newsec#1{\global\advance\secno by1\message{(\the\secno. #1)}
\global\subsecno=0\eqnres@t\noindent{\bf\the\secno. #1}
\writetoca{{\secsym} {#1}}\par\nobreak\medskip\nobreak}
\def\eqnres@t{\xdef\secsym{\the\secno.}\global\meqno=1\bigbreak\bigskip}
\def\sequentialequations{\def\eqnres@t{\bigbreak}}\xdef\secsym{}
\global\newcount\subsecno \global\subsecno=0
\def\subsec#1{\global\advance\subsecno by1\message{(\secsym\the\subsecno. #1)}
\ifnum\lastpenalty>9000\else\bigbreak\fi
\noindent{\it\secsym\the\subsecno. #1}\writetoca{\string\quad
{\secsym\the\subsecno.} {#1}}\par\nobreak\medskip\nobreak}
\def\appendix#1#2{\global\meqno=1\global\subsecno=0\xdef\secsym{\hbox{#1.}}
\bigbreak\bigskip\noindent{\bf Appendix #1. #2}\message{(#1. #2)}
\writetoca{Appendix {#1.} {#2}}\par\nobreak\medskip\nobreak}
%
%
\def\eqnn#1{\xdef #1{(\secsym\the\meqno)}\writedef{#1\leftbracket#1}%
\global\advance\meqno by1\wrlabeL#1}
\def\eqna#1{\xdef #1##1{\hbox{$(\secsym\the\meqno##1)$}}
\writedef{#1\numbersign1\leftbracket#1{\numbersign1}}%
\global\advance\meqno by1\wrlabeL{#1$\{\}$}}
\def\eqn#1#2{\xdef #1{(\secsym\the\meqno)}\writedef{#1\leftbracket#1}%
\global\advance\meqno by1$$#2\eqno#1\eqlabeL#1$$}
%
\newskip\footskip\footskip14pt plus 1pt minus 1pt 
\def\footnotefont{\ninepoint}\def\f@t#1{\footnotefont #1\@foot}
\def\f@@t{\baselineskip\footskip\bgroup\footnotefont\aftergroup\@foot\let\next}
\setbox\strutbox=\hbox{\vrule height9.5pt depth4.5pt width0pt}
\global\newcount\ftno \global\ftno=0
\def\foot{\global\advance\ftno by1\footnote{$^{\the\ftno}$}}
%
\newwrite\ftfile
\def\footend{\def\foot{\global\advance\ftno by1\chardef\wfile=\ftfile
$^{\the\ftno}$\ifnum\ftno=1\immediate\openout\ftfile=foots.tmp\fi%
\immediate\write\ftfile{\noexpand\smallskip%
\noexpand\item{f\the\ftno:\ }\pctsign}\findarg}%
\def\footatend{\vfill\eject\immediate\closeout\ftfile{\parindent=20pt
\centerline{\bf Footnotes}\nobreak\bigskip\input foots.tmp }}}
\def\footatend{}
%
%
\global\newcount\refno \global\refno=1
\newwrite\rfile
\def\ref{[\the\refno]\nref}
\def\nref#1{\xdef#1{[\the\refno]}\writedef{#1\leftbracket#1}%
\ifnum\refno=1\immediate\openout\rfile=refs.tmp\fi
\global\advance\refno by1\chardef\wfile=\rfile\immediate
\write\rfile{\noexpand\item{#1\ }\reflabeL{#1\hskip.31in}\pctsign}\findarg}
\def\findarg#1#{\begingroup\obeylines\newlinechar=`\^^M\pass@rg}
{\obeylines\gdef\pass@rg#1{\writ@line\relax #1^^M\hbox{}^^M}%
\gdef\writ@line#1^^M{\expandafter\toks0\expandafter{\striprel@x #1}%
\edef\next{\the\toks0}\ifx\next\em@rk\let\next=\endgroup\else\ifx\next\empty%
\else\immediate\write\wfile{\the\toks0}\fi\let\next=\writ@line\fi\next\relax}}
\def\striprel@x#1{} \def\em@rk{\hbox{}}
\def\lref{\begingroup\obeylines\lr@f}
\def\lr@f#1#2{\gdef#1{\ref#1{#2}}\endgroup\unskip}

\def\addref#1{\immediate\write\rfile{\noexpand\item{}#1}} 
\def\startrefs#1{\immediate\openout\rfile=refs.tmp\refno=#1}
\def\xref{\expandafter\xr@f}\def\xr@f[#1]{#1}
\def\refs#1{\count255=1[\r@fs #1{\hbox{}}]}
\def\r@fs#1{\ifx\und@fined#1\message{reflabel \string#1 is undefined.}%
\nref#1{need to supply reference \string#1.}\fi%
\vphantom{\hphantom{#1}}\edef\next{#1}\ifx\next\em@rk\def\next{}%
\else\ifx\next#1\ifodd\count255\relax\xref#1\count255=0\fi%
\else#1\count255=1\fi\let\next=\r@fs\fi\next}
%

%
\newwrite\ffile\global\newcount\figno \global\figno=1
\def\fig{fig.~\the\figno\nfig}
\def\nfig#1{\xdef#1{fig.~\the\figno}%
\writedef{#1\leftbracket fig.\noexpand~\the\figno}%
\ifnum\figno=1\immediate\openout\ffile=figs.tmp\fi\chardef\wfile=\ffile%
\immediate\write\ffile{\noexpand\medskip\noexpand\item{Fig.\ \the\figno. }
\reflabeL{#1\hskip.55in}\pctsign}\global\advance\figno by1\findarg}
\def\xfig{\expandafter\xf@g}\def\xf@g fig.\penalty\@M\ {}
\def\figs#1{figs.~\f@gs #1{\hbox{}}}
\def\f@gs#1{\edef\next{#1}\ifx\next\em@rk\def\next{}\else
\ifx\next#1\xfig #1\else#1\fi\let\next=\f@gs\fi\next}
\newwrite\lfile
{\escapechar-1\xdef\pctsign{\string\%}\xdef\leftbracket{\string\{}
\xdef\rightbracket{\string\}}\xdef\numbersign{\string\#}}

\def\writestop{\def\writestoppt{\immediate\write\lfile{\string\pageno%
\the\pageno\string\startrefs\leftbracket\the\refno\rightbracket%
\string\def\string\secsym\leftbracket\secsym\rightbracket%
\string\secno\the\secno\string\meqno\the\meqno}\immediate\closeout\lfile}}
\def\writestoppt{}\def\writedef#1{}
\def\seclab#1{\xdef #1{\the\secno}\writedef{#1\leftbracket#1}\wrlabeL{#1=#1}}
\def\subseclab#1{\xdef #1{\secsym\the\subsecno}%
\writedef{#1\leftbracket#1}\wrlabeL{#1=#1}}
\newwrite\tfile \def\writetoca#1{}
\def\leaderfill{\leaders\hbox to 1em{\hss.\hss}\hfill}
\def\writetoc{\immediate\openout\tfile=toc.tmp
   \def\writetoca##1{{\edef\next{\write\tfile{\noindent ##1
   \string\leaderfill {\noexpand\number\pageno} \par}}\next}}}

%
\catcode`\@=12 
%
\edef\tfontsize{\ifx\answ\bigans scaled\magstep3\else scaled\magstep4\fi}
\font\titlerm=cmr10 \tfontsize \font\titlerms=cmr7 \tfontsize
\font\titlermss=cmr5 \tfontsize \font\titlei=cmmi10 \tfontsize
\font\titleis=cmmi7 \tfontsize \font\titleiss=cmmi5 \tfontsize
\font\titlesy=cmsy10 \tfontsize \font\titlesys=cmsy7 \tfontsize
\font\titlesyss=cmsy5 \tfontsize \font\titleit=cmti10 \tfontsize
\skewchar\titlei='177 \skewchar\titleis='177 \skewchar\titleiss='177
\skewchar\titlesy='60 \skewchar\titlesys='60 \skewchar\titlesyss='60
\def\titlefont{\def\rm{\fam0\titlerm}
\textfont0=\titlerm \scriptfont0=\titlerms \scriptscriptfont0=\titlermss
\textfont1=\titlei \scriptfont1=\titleis \scriptscriptfont1=\titleiss
\textfont2=\titlesy \scriptfont2=\titlesys \scriptscriptfont2=\titlesyss
\textfont\itfam=\titleit \def\it{\fam\itfam\titleit}\rm}
 \ifx\answ\bigans\else scaled\magstep1\fi
\ifx\answ\bigans\def\abstractfont{\tenpoint}\else
\font\abssl=cmsl10 scaled \magstep1
\font\absrm=cmr10 scaled\magstep1 \font\absrms=cmr7 scaled\magstep1
\font\absrmss=cmr5 scaled\magstep1 \font\absi=cmmi10 scaled\magstep1
\font\absis=cmmi7 scaled\magstep1 \font\absiss=cmmi5 scaled\magstep1
\font\abssy=cmsy10 scaled\magstep1 \font\abssys=cmsy7 scaled\magstep1
\font\abssyss=cmsy5 scaled\magstep1 \font\absbf=cmbx10 scaled\magstep1
\skewchar\absi='177 \skewchar\absis='177 \skewchar\absiss='177
\skewchar\abssy='60 \skewchar\abssys='60 \skewchar\abssyss='60
\def\abstractfont{\def\rm{\fam0\absrm}
\textfont0=\absrm \scriptfont0=\absrms \scriptscriptfont0=\absrmss
\textfont1=\absi \scriptfont1=\absis \scriptscriptfont1=\absiss
\textfont2=\abssy \scriptfont2=\abssys \scriptscriptfont2=\abssyss
\textfont\itfam=\bigit \def\it{\fam\itfam\bigit}\def\footnotefont{\tenpoint}%
\textfont\slfam=\abssl \def\sl{\fam\slfam\abssl}%
\textfont\bffam=\absbf \def\bf{\fam\bffam\absbf}\rm}\fi
\def\tenpoint{\def\rm{\fam0\tenrm}
\textfont0=\tenrm \scriptfont0=\sevenrm \scriptscriptfont0=\fiverm
\textfont1=\teni  \scriptfont1=\seveni  \scriptscriptfont1=\fivei
\textfont2=\tensy \scriptfont2=\sevensy \scriptscriptfont2=\fivesy
\textfont\itfam=\tenit \def\it{\fam\itfam\tenit}\def\footnotefont{\ninepoint}%
\textfont\bffam=\tenbf \def\bf{\fam\bffam\tenbf}\def\sl{\fam\slfam\tensl}\rm}
\font\ninerm=cmr9 \font\sixrm=cmr6 \font\ninei=cmmi9 \font\sixi=cmmi6
\font\ninesy=cmsy9 \font\sixsy=cmsy6 \font\ninebf=cmbx9
\font\nineit=cmti9 \font\ninesl=cmsl9 \skewchar\ninei='177
\skewchar\sixi='177 \skewchar\ninesy='60 \skewchar\sixsy='60
\def\ninepoint{\def\rm{\fam0\ninerm}
\textfont0=\ninerm \scriptfont0=\sixrm \scriptscriptfont0=\fiverm
\textfont1=\ninei \scriptfont1=\sixi \scriptscriptfont1=\fivei
\textfont2=\ninesy \scriptfont2=\sixsy \scriptscriptfont2=\fivesy
\textfont\itfam=\ninei \def\it{\fam\itfam\nineit}\def\sl{\fam\slfam\ninesl}%
\textfont\bffam=\ninebf \def\bf{\fam\bffam\ninebf}\rm}
%
%

\hyphenation{anom-aly anom-alies coun-ter-term coun-ter-terms}
\def\inv{^{\raise.15ex\hbox{${\scriptscriptstyle -}$}\kern-.05em 1}}

\def\Dsl{\,\raise.15ex\hbox{/}\mkern-13.5mu D} 
\def\dsl{\raise.15ex\hbox{/}\kern-.57em\partial}

\font\bigit=cmti10 scaled \magstep1
\def\lspace{\ifx\answ\bigans{}\else\qquad\fi}
\def\lbspace{\ifx\answ\bigans{}\else\hskip-.2in\fi} 
\def\boxeqn#1{\vcenter{\vbox{\hrule\hbox{\vrule\kern3pt\vbox{\kern3pt
    \hbox{${\displaystyle #1}$}\kern3pt}\kern3pt\vrule}\hrule}}}
\def\mbox#1#2{\vcenter{\hrule \hbox{\vrule height#2in
        \kern#1in \vrule} \hrule}}  
%

\def\darr#1{\raise1.5ex\hbox{$\leftrightarrow$}\mkern-16.5mu #1}

\def\roughly#1{\raise.3ex\hbox{$#1$\kern-.75em\lower1ex\hbox{$\sim$}}}

\let\includefigures=\iftrue
\let\useblackboard=\iftrue
\newfam\black

\includefigures
\message{If you do not have epsf.tex (to include figures),}
\message{change the option at the top of the tex file.}
\input epsf
\def\figin{\epsfcheck\figin}\def\figins{\epsfcheck\figins}
\def\epsfcheck{\ifx\epsfbox\UnDeFiNeD
\message{(NO epsf.tex, FIGURES WILL BE IGNORED)}
\gdef\figin##1{\vskip2in}\gdef\figins##1{\hskip.5in}
\else\message{(FIGURES WILL BE INCLUDED)}%
\gdef\figin##1{##1}\gdef\figins##1{##1}\fi}
\def\DefWarn#1{}
\def\figinsert{\goodbreak\midinsert}
\def\ifig#1#2#3{\DefWarn#1\xdef#1{fig.~\the\figno}
\writedef{#1\leftbracket fig.\noexpand~\the\figno}%
\figinsert\figin{\centerline{#3}}\medskip\centerline{\vbox{
\baselineskip12pt\advance\hsize by -1truein
\noindent\footnotefont{\bf Fig.~\the\figno:} #2}}
\endinsert\global\advance\figno by1}
\else
\def\ifig#1#2#3{\xdef#1{fig.~\the\figno}
\writedef{#1\leftbracket fig.\noexpand~\the\figno}%
\global\advance\figno by1} \fi

\def\id{{1 \kern-.28em {\rm l}}}

\def\K3{{\bf K3}}
\def\journal#1&#2(#3){\unskip, \sl #1\ \bf #2 \rm(19#3) }
\def\andjournal#1&#2(#3){\sl #1~\bf #2 \rm (19#3) }

\def\hat{\widehat}

\def\tilde{\widetilde}

\def\frac#1#2{{#1\over#2}}

\def\inbar{\,\vrule height1.5ex width.4pt depth0pt}
\def\IC{\relax\hbox{$\inbar\kern-.3em{\rm C}$}}
\def\IR{\relax{\rm I\kern-.18em R}}
\def\IP{\relax{\rm I\kern-.18em P}}

%
%

%
\catcode`\@=11
\def\slash#1{\mathord{\mathpalette\c@ncel{#1}}}
\overfullrule=0pt

\def\DD{{\cal D}}

\def\LL{{\cal L}}

\def\OO{{\cal O}}

\def\RR{{\cal R}}
\def\SS{{\cal S}}

\def\ZZ{{\cal Z}}

\def\underrel#1\over#2{\mathrel{\mathop{\kern\z@#1}\limits_{#2}}}

\catcode`\@=12


%

\def \sinh{{\rm sinh}}
\def \cosh{{\rm cosh}}

\def\exp{{\rm exp}}


\def\p{{\partial}}

\def\ra{{\rightarrow}}

\def\LL{{\cal L}}

\lref\KulaxiziGY{
  M.~Kulaxizi, A.~Parnachev and K.~Schalm,
  ``On Holographic Entanglement Entropy of Charged Matter,''
[arXiv:1208.2937 [hep-th]].
}

\lref\CasiniEI{
  H.~Casini and M.~Huerta,
  ``On the RG running of the entanglement entropy of a circle,''
Phys.\ Rev.\ D {\bf 85}, 125016 (2012).
[arXiv:1202.5650 [hep-th]].
}

\lref\PakmanUI{
  A.~Pakman and A.~Parnachev,
  ``Topological Entanglement Entropy and Holography,''
JHEP {\bf 0807}, 097 (2008).
[arXiv:0805.1891 [hep-th]].
}

\lref\deBoerWK{
  J.~de Boer, M.~Kulaxizi and A.~Parnachev,
  ``Holographic Entanglement Entropy in Lovelock Gravities,''
JHEP {\bf 1107}, 109 (2011).
[arXiv:1101.5781 [hep-th]].
}

\lref\HungXB{
  L.~-Y.~Hung, R.~C.~Myers and M.~Smolkin,
  ``On Holographic Entanglement Entropy and Higher Curvature Gravity,''
JHEP {\bf 1104}, 025 (2011).
[arXiv:1101.5813 [hep-th]].
}

\lref\BatellME{
  B.~Batell, T.~Gherghetta and D.~Sword,
  ``The Soft-Wall Standard Model,''
Phys.\ Rev.\ D {\bf 78}, 116011 (2008).
[arXiv:0808.3977 [hep-ph]].
}

\lref\CasiniKV{
  H.~Casini, M.~Huerta and R.~C.~Myers,
  ``Towards a derivation of holographic entanglement entropy,''
JHEP {\bf 1105}, 036 (2011).
[arXiv:1102.0440 [hep-th]].
}
\lref\LevinZZ{
  M.~Levin and X.~G.~Wen,
  ``Detecting Topological Order in a Ground State Wave Function,''
Phys.\ Rev.\ Lett.\  {\bf 96}, 110405 (2006).
}
\lref\KitaevDM{
  A.~Kitaev and J.~Preskill,
  ``Topological entanglement entropy,''
Phys.\ Rev.\ Lett.\  {\bf 96}, 110404 (2006).
[hep-th/0510092].
}
\lref\RyuBV{
  S.~Ryu and T.~Takayanagi,
  ``Holographic derivation of entanglement entropy from AdS/CFT,''
Phys.\ Rev.\ Lett.\  {\bf 96}, 181602 (2006).
[hep-th/0603001],
S.~Ryu and T.~Takayanagi,
  ``Aspects of Holographic Entanglement Entropy,''
JHEP {\bf 0608}, 045 (2006).
[hep-th/0605073].
}
\lref\HeadrickKM{
  M.~Headrick and T.~Takayanagi,
  ``A Holographic proof of the strong subadditivity of entanglement entropy,''
Phys.\ Rev.\ D {\bf 76}, 106013 (2007).
[arXiv:0704.3719 [hep-th]].
}
\lref\LewkowyczNQA{
  A.~Lewkowycz and J.~Maldacena,
  ``Generalized gravitational entropy,''
JHEP {\bf 1308}, 090 (2013).
[arXiv:1304.4926 [hep-th]].
}
\lref\DuffWM{
  M.~J.~Duff,
  ``Twenty years of the Weyl anomaly,''
Class.\ Quant.\ Grav.\  {\bf 11}, 1387 (1994).
[hep-th/9308075].
}
\lref\BuchelVZ{
  A.~Buchel, R.~C.~Myers and A.~Sinha,
  ``Beyond eta/s = 1/4 pi,''
JHEP {\bf 0903}, 084 (2009).
[arXiv:0812.2521 [hep-th]].
}
\lref\BriganteNU{
  M.~Brigante, H.~Liu, R.~C.~Myers, S.~Shenker and S.~Yaida,
  ``Viscosity Bound Violation in Higher Derivative Gravity,''
Phys.\ Rev.\ D {\bf 77}, 126006 (2008).
[arXiv:0712.0805 [hep-th]].
}
\lref\BuchelSK{
  A.~Buchel, J.~Escobedo, R.~C.~Myers, M.~F.~Paulos, A.~Sinha and M.~Smolkin,
  ``Holographic GB gravity in arbitrary dimensions,''
JHEP {\bf 1003}, 111 (2010).
[arXiv:0911.4257 [hep-th]].
}
\lref\deBoerPN{
  J.~de Boer, M.~Kulaxizi and A.~Parnachev,
  ``AdS(7)/CFT(6), Gauss-Bonnet Gravity, and Viscosity Bound,''
JHEP {\bf 1003}, 087 (2010).
[arXiv:0910.5347 [hep-th]].
}
\lref\MyersXS{
  R.~C.~Myers and A.~Sinha,
  ``Seeing a c-theorem with holography,''
Phys.\ Rev.\ D {\bf 82}, 046006 (2010).
[arXiv:1006.1263 [hep-th]].
  R.~C.~Myers and A.~Sinha,
  ``Holographic c-theorems in arbitrary dimensions,''
JHEP {\bf 1101}, 125 (2011).
[arXiv:1011.5819 [hep-th]].
}
\lref\FursaevIH{
  D.~V.~Fursaev,
  ``Proof of the holographic formula for entanglement entropy,''
JHEP {\bf 0609}, 018 (2006).
[hep-th/0606184].
}
\lref\DongQOA{
  X.~Dong,
  ``Holographic Entanglement Entropy for General Higher Derivative Gravity,''
JHEP {\bf 1401}, 044 (2014).
[arXiv:1310.5713 [hep-th], arXiv:1310.5713].
}
\lref\CampsZUA{
  J.~Camps,
  ``Generalized entropy and higher derivative Gravity,''
JHEP {\bf 1403}, 070 (2014).
[arXiv:1310.6659 [hep-th]].
}
\lref\LiuEEA{
  H.~Liu and M.~Mezei,
  ``A Refinement of entanglement entropy and the number of degrees of freedom,''
JHEP {\bf 1304}, 162 (2013).
[arXiv:1202.2070 [hep-th]].
}
\lref\LiuUNA{
  H.~Liu and M.~Mezei,
JHEP {\bf 1401}, 098 (2014).
[arXiv:1309.6935 [hep-th]].
}
\lref\KlebanovYF{
  I.~R.~Klebanov, T.~Nishioka, S.~S.~Pufu and B.~R.~Safdi,
  ``On Shape Dependence and RG Flow of Entanglement Entropy,''
JHEP {\bf 1207}, 001 (2012).
[arXiv:1204.4160 [hep-th]].
}
\lref\AllaisATA{
  A.~Allais and M.~Mezei,
  ``Some results on the shape dependence of entanglement and R\'enyi entropies,''
[arXiv:1407.7249 [hep-th]].
}
\lref\HubenyRY{
  V.~E.~Hubeny,
  ``Extremal surfaces as bulk probes in AdS/CFT,''
JHEP {\bf 1207}, 093 (2012).
[arXiv:1203.1044 [hep-th]].
}
\lref\KarchPV{
  A.~Karch, E.~Katz, D.~T.~Son and M.~A.~Stephanov,
  ``Linear confinement and AdS/QCD,''
Phys.\ Rev.\ D {\bf 74}, 015005 (2006).
[hep-ph/0602229].
}
\lref\CacciapagliaNS{
  G.~Cacciapaglia, G.~Marandella and J.~Terning,
  ``The AdS/CFT/Unparticle Correspondence,''
JHEP {\bf 0902}, 049 (2009).
[arXiv:0804.0424 [hep-ph]].
}
\lref\KlebanovWS{
  I.~R.~Klebanov, D.~Kutasov and A.~Murugan,
  ``Entanglement as a probe of confinement,''
Nucl.\ Phys.\ B {\bf 796}, 274 (2008).
[arXiv:0709.2140 [hep-th]].
}
\lref\KolNQA{
  U.~Kol, C.~Nunez, D.~Schofield, J.~Sonnenschein and M.~Warschawski,
  ``Confinement, Phase Transitions and non-Locality in the Entanglement Entropy,''
JHEP {\bf 1406}, 005 (2014).
[arXiv:1403.2721 [hep-th]].
}
\lref\HerzogRA{
  C.~P.~Herzog,
  ``A Holographic Prediction of the Deconfinement Temperature,''
Phys.\ Rev.\ Lett.\  {\bf 98}, 091601 (2007).
[hep-th/0608151].
}
\lref\EmparanPM{
  R.~Emparan, C.~V.~Johnson and R.~C.~Myers,
  ``Surface terms as counterterms in the AdS / CFT correspondence,''
Phys.\ Rev.\ D {\bf 60}, 104001 (1999).
[hep-th/9903238].
}
\lref\SkenderisMM{
  K.~Skenderis and P.~K.~Townsend,
  ``Gravitational stability and renormalization group flow,''
Phys.\ Lett.\ B {\bf 468}, 46 (1999).
[hep-th/9909070].
}
\lref\BatellZM{
  B.~Batell and T.~Gherghetta,
  ``Dynamical Soft-Wall AdS/QCD,''
Phys.\ Rev.\ D {\bf 78}, 026002 (2008).
[arXiv:0801.4383 [hep-ph]].
}
\lref\KelleyVZ{
  T.~M.~Kelley,
  ``The Thermodynamics of a 5D Gravity-Dilaton-Tachyon Solution,''
[arXiv:1107.0931 [hep-ph]].
}
\lref\EmparanGF{
  R.~Emparan,
  ``AdS / CFT duals of topological black holes and the entropy of zero energy states,''
JHEP {\bf 9906}, 036 (1999).
[hep-th/9906040].
}
\lref\IyerYS{
  V.~Iyer and R.~M.~Wald,
  ``Some properties of Noether charge and a proposal for dynamical black hole entropy,''
Phys.\ Rev.\ D {\bf 50}, 846 (1994).
[gr-qc/9403028].
 V.~Iyer and R.~M.~Wald,
  ``A Comparison of Noether charge and Euclidean methods for computing the entropy of stationary black holes,''
Phys.\ Rev.\ D {\bf 52}, 4430 (1995).
[gr-qc/9503052].
 R.~M.~Wald and V.~Iyer,
  ``Trapped surfaces in the Schwarzschild geometry and cosmic censorship,''
Phys.\ Rev.\ D {\bf 44}, 3719 (1991).
}
\lref\HeadrickZDA{
  M.~Headrick,
  ``General properties of holographic entanglement entropy,''
JHEP {\bf 1403}, 085 (2014).
[arXiv:1312.6717 [hep-th]].
}
\lref\WehrlZZ{
  A.~Wehrl,
  ``General properties of entropy,''
Rev.\ Mod.\ Phys.\  {\bf 50}, 221 (1978).
}
\lref\DongFT{
  S.~Dong, E.~Fradkin, R.~G.~Leigh and S.~Nowling,
  ``Topological Entanglement Entropy in Chern-Simons Theories and Quantum Hall Fluids,''
JHEP {\bf 0805}, 016 (2008).
[arXiv:0802.3231 [hep-th]].
}
\lref\HirataJX{
  T.~Hirata and T.~Takayanagi,
  ``AdS/CFT and strong subadditivity of entanglement entropy,''
JHEP {\bf 0702}, 042 (2007).
[hep-th/0608213].
}
\lref\SarkarXP{
  S.~Sarkar and A.~C.~Wall,
  ``Second Law Violations in Lovelock Gravity for Black Hole Mergers,''
Phys.\ Rev.\ D {\bf 83}, 124048 (2011).
[arXiv:1011.4988 [gr-qc]].
}
\lref\ClunanTB{
  T.~Clunan, S.~F.~Ross and D.~J.~Smith,
  ``On Gauss-Bonnet black hole entropy,''
Class.\ Quant.\ Grav.\  {\bf 21}, 3447 (2004).
[gr-qc/0402044].
}
\lref\KovtunEV{
  P.~K.~Kovtun and A.~O.~Starinets,
  ``Quasinormal modes and holography,''
Phys.\ Rev.\ D {\bf 72}, 086009 (2005).
[hep-th/0506184].
}
\lref\EdalatiPN{
  M.~Edalati, J.~I.~Jottar and R.~G.~Leigh,
  ``Holography and the sound of criticality,''
JHEP {\bf 1010}, 058 (2010).
[arXiv:1005.4075 [hep-th]].
}
\lref\DavisonBXA{
  R.~A.~Davison and A.~Parnachev,
  ``Hydrodynamics of cold holographic matter,''
JHEP {\bf 1306}, 100 (2013).
[arXiv:1303.6334 [hep-th]].
}
\lref\LippertJMA{
  M.~Lippert, R.~Meyer and A.~Taliotis,
  ``A holographic model for the fractional quantum Hall effect,''
[arXiv:1409.1369 [hep-th]].
}
\lref\ErdmengerRCA{
  J.~Erdmenger, D.~W.~Pang and H.~Zeller,
  ``Holographic entanglement entropy of semi-local quantum liquids,''
JHEP {\bf 1402}, 016 (2014).
[arXiv:1311.1217 [hep-th]].
}
\lref\CveticBK{
  M.~Cvetic, S.~Nojiri and S.~D.~Odintsov,
  ``Black hole thermodynamics and negative entropy in de Sitter and anti-de Sitter Einstein-Gauss-Bonnet gravity,''
Nucl.\ Phys.\ B {\bf 628}, 295 (2002).
[hep-th/0112045].
}
\lref\ForcellaDWA{
  D.~Forcella, A.~Mezzalira and D.~Musso,
  ``Electromagnetic response of strongly coupled plasmas,''
[arXiv:1404.4048 [hep-th]].
}
\lref\AminneborgIZ{
  S.~Aminneborg, I.~Bengtsson, S.~Holst and P.~Peldan,
  ``Making anti-de Sitter black holes,''
Class.\ Quant.\ Grav.\  {\bf 13}, 2707 (1996).
[gr-qc/9604005].
}
\lref\WittenHF{
  E.~Witten,
  ``Quantum Field Theory and the Jones Polynomial,''
Commun.\ Math.\ Phys.\  {\bf 121}, 351 (1989)..
}
\lref\MooreYH{
  G.~W.~Moore and N.~Seiberg,
  ``Taming the Conformal Zoo,''
Phys.\ Lett.\ B {\bf 220}, 422 (1989)..
}
\lref\VerlindeSN{
  E.~P.~Verlinde,
  ``Fusion Rules and Modular Transformations in 2D Conformal Field Theory,''
Nucl.\ Phys.\ B {\bf 300}, 360 (1988)..
}
\lref\BlauTV{
  M.~Blau and G.~Thompson,
  ``Derivation of the Verlinde formula from Chern-Simons theory and the G/G model,''
Nucl.\ Phys.\ B {\bf 408}, 345 (1993).
[hep-th/9305010].
}
\lref\DiFrancescoNK{
  P.~Di Francesco, P.~Mathieu and D.~Senechal,
  ``Conformal field theory,''
New York, USA: Springer (1997) 890 p.
}
\lref\NayakZZA{
  C.~Nayak, S.~H.~Simon, A.~Stern, M.~Freedman and S.~Das Sarma,
  ``Non-Abelian anyons and topological quantum computation,''
Rev.\ Mod.\ Phys.\  {\bf 80}, 1083 (2008).
[arXiv:0707.1889 [cond-mat.str-el]]
}
\lref\WenZG{
  X.~G.~Wen,
  ``Vacuum Degeneracy of Chiral Spin States in Compactified Space,''
Phys.\ Rev.\ B {\bf 40}, 7387 (1989)..
}
\lref\FendleyGR{
  P.~Fendley, M.~P.~A.~Fisher and C.~Nayak,
  ``Topological entanglement entropy from the holographic partition function,''
J.\ Statist.\ Phys.\  {\bf 126}, 1111 (2007).
[cond-mat/0609072 [cond-mat.stat-mech]].
}
\lref\WittenWE{
  E.~Witten,
  ``On quantum gauge theories in two-dimensions,''
Commun.\ Math.\ Phys.\  {\bf 141}, 153 (1991)..
}
\lref\CamperiDK{
  M.~Camperi, F.~Levstein and G.~Zemba,
  ``The Large $N$ Limit of {Chern-Simons} Gauge Theory,''
Phys.\ Lett.\ B {\bf 247}, 549 (1990)..
}
\lref\IsidroEM{
  J.~M.~Isidro, J.~P.~Nunes and H.~J.~Schnitzer,
  ``BF theories and group level duality,''
Nucl.\ Phys.\ B {\bf 465}, 315 (1996).
[hep-th/9510064].
}
\lref\BayntunNX{
  A.~Bayntun, C.~P.~Burgess, B.~P.~Dolan and S.~S.~Lee,
  ``AdS/QHE: Towards a Holographic Description of Quantum Hall Experiments,''
New J.\ Phys.\  {\bf 13}, 035012 (2011).
[arXiv:1008.1917 [hep-th]].
}
\lref\BergmanGM{
  O.~Bergman, N.~Jokela, G.~Lifschytz and M.~Lippert,
  ``Quantum Hall Effect in a Holographic Model,''
JHEP {\bf 1010}, 063 (2010).
[arXiv:1003.4965 [hep-th]].
}
\lref\KristjansenNY{
  C.~Kristjansen and G.~W.~Semenoff,
  ``Giant D5 Brane Holographic Hall State,''
JHEP {\bf 1306}, 048 (2013).
[arXiv:1212.5609 [hep-th]].
}
\lref\FujitaKW{
  M.~Fujita, W.~Li, S.~Ryu and T.~Takayanagi,
  ``Fractional Quantum Hall Effect via Holography: Chern-Simons, Edge States, and Hierarchy,''
JHEP {\bf 0906}, 066 (2009).
[arXiv:0901.0924 [hep-th]].
}
\lref\BanerjeeGH{
  S.~Banerjee, S.~Hellerman, J.~Maltz and S.~H.~Shenker,
  ``Light States in Chern-Simons Theory Coupled to Fundamental Matter,''
JHEP {\bf 1303}, 097 (2013).
[arXiv:1207.4195 [hep-th]].
}
\lref\CamanhoAPA{
  X.~O.~Camanho, J.~D.~Edelstein, J.~Maldacena and A.~Zhiboedov,
  ``Causality Constraints on Corrections to the Graviton Three-Point Coupling,''
[arXiv:1407.5597 [hep-th]].
}
\lref\SolodukhinDH{
  S.~N.~Solodukhin,
  ``Entanglement entropy, conformal invariance and extrinsic geometry,''
Phys.\ Lett.\ B {\bf 665}, 305 (2008).
[arXiv:0802.3117 [hep-th]].
}
\lref\MooreUZ{
  G.~W.~Moore and N.~Seiberg,
  ``Polynomial Equations for Rational Conformal Field Theories,''
Phys.\ Lett.\ B {\bf 212}, 451 (1988) ; 
  G.~W.~Moore and N.~Seiberg,
  ``Naturality in Conformal Field Theory,''
Nucl.\ Phys.\ B {\bf 313}, 16 (1989)..
}
\lref\ElitzurNR{
  S.~Elitzur, G.~W.~Moore, A.~Schwimmer and N.~Seiberg,
  ``Remarks on the Canonical Quantization of the Chern-Simons-Witten Theory,''
Nucl.\ Phys.\ B {\bf 326}, 108 (1989)..
}
\lref\DijkgraafTF{
  R.~Dijkgraaf and E.~P.~Verlinde,
  ``Modular Invariance and the Fusion Algebra,''
Nucl.\ Phys.\ Proc.\ Suppl.\  {\bf 5B}, 87 (1988)..
}
\lref\MarinoUF{
  M.~Marino,
  ``Chern-Simons theory and topological strings,''
Rev.\ Mod.\ Phys.\  {\bf 77}, 675 (2005).
[hep-th/0406005].
}
\lref\FalkowskiFZ{
  A.~Falkowski and M.~Perez-Victoria,
  ``Electroweak Breaking on a Soft Wall,''
JHEP {\bf 0812}, 107 (2008).
[arXiv:0806.1737 [hep-ph]].
}

\lref\DavisNV{
  J.~L.~Davis, P.~Kraus and A.~Shah,
  ``Gravity Dual of a Quantum Hall Plateau Transition,''
JHEP {\bf 0811}, 020 (2008).
[arXiv:0809.1876 [hep-th]].
}
\lref\BelhajIW{
  A.~Belhaj, N.~E.~Fahssi, E.~H.~Saidi and A.~Segui,
  ``Embedding Fractional Quantum Hall Solitons in M-theory Compactifications,''
Int.\ J.\ Geom.\ Meth.\ Mod.\ Phys.\  {\bf 8}, 1507 (2011).
[arXiv:1007.4485 [hep-th]].
}
\lref\GubankovaRC{
  E.~Gubankova, J.~Brill, M.~Cubrovic, K.~Schalm, P.~Schijven and J.~Zaanen,
  ``Holographic fermions in external magnetic fields,''
Phys.\ Rev.\ D {\bf 84}, 106003 (2011).
[arXiv:1011.4051 [hep-th]].
}
\lref\JokelaEB{
  N.~Jokela, M.~Jarvinen and M.~Lippert,
  ``A holographic quantum Hall model at integer filling,''
JHEP {\bf 1105}, 101 (2011).
[arXiv:1101.3329 [hep-th]].
}
\lref\BlakeTP{
  M.~Blake, S.~Bolognesi, D.~Tong and K.~Wong,
  ``Holographic Dual of the Lowest Landau Level,''
JHEP {\bf 1212}, 039 (2012).
[arXiv:1208.5771 [hep-th]].
}
\lref\FujitaFP{
  M.~Fujita, M.~Kaminski and A.~Karch,
  ``SL(2,Z) Action on AdS/BCFT and Hall Conductivities,''
JHEP {\bf 1207}, 150 (2012).
[arXiv:1204.0012 [hep-th]].
}
\lref\AlanenCN{
  J.~Alanen, E.~Keski-Vakkuri, P.~Kraus and V.~Suur-Uski,
  ``AC Transport at Holographic Quantum Hall Transitions,''
JHEP {\bf 0911}, 014 (2009).
[arXiv:0905.4538 [hep-th]].
}
\lref\NishiokaGR{
  T.~Nishioka and T.~Takayanagi,
  ``AdS Bubbles, Entropy and Closed String Tachyons,''
JHEP {\bf 0701}, 090 (2007).
[hep-th/0611035].
}
\lref\CaiII{
  R.~G.~Cai, J.~Y.~Ji and K.~S.~Soh,
  ``Topological dilaton black holes,''
Phys.\ Rev.\ D {\bf 57}, 6547 (1998).
[gr-qc/9708063].
}
\lref\McGoughGKA{
  L.~McGough and H.~Verlinde,
  ``Bekenstein-Hawking Entropy as Topological Entanglement Entropy,''
JHEP {\bf 1311}, 208 (2013).
[arXiv:1308.2342 [hep-th]].
}
\lref\LiHP{
  D.~Li, S.~He, M.~Huang and Q.~S.~Yan,
  ``Thermodynamics of deformed AdS$_5$ model with a positive/negative quadratic correction in graviton-dilaton system,''
JHEP {\bf 1109}, 041 (2011).
[arXiv:1103.5389 [hep-th]].
}
\lref\MiaoNXA{
  R.~X.~Miao and W.~z.~Guo,
  ``Holographic Entanglement Entropy for the Most General Higher Derivative Gravity,''
[arXiv:1411.5579 [hep-th]].
}
\lref\AmiGSA{
  O.~Ben-Ami, D.~Carmi and J.~Sonnenschein,
  ``Holographic Entanglement Entropy of Multiple Strips,''
JHEP {\bf 1411}, 144 (2014).
[arXiv:1409.6305 [hep-th]].
}
\lref\BhattacharyyaGRA{
  A.~Bhattacharyya, M.~Sharma and A.~Sinha,
  ``On generalized gravitational entropy, squashed cones and holography,''
JHEP {\bf 1401}, 021 (2014).
[arXiv:1308.5748 [hep-th]].
}
\lref\BhattacharyyaYGA{
  A.~Bhattacharyya and M.~Sharma,
  ``On entanglement entropy functionals in higher derivative gravity theories,''
JHEP {\bf 1410}, 130 (2014).
[arXiv:1405.3511 [hep-th]].
}
\lref\JacobsonXS{
  T.~Jacobson and R.~C.~Myers,
  ``Black hole entropy and higher curvature interactions,''
Phys.\ Rev.\ Lett.\  {\bf 70}, 3684 (1993).
[hep-th/9305016].
}
\lref\LikoVI{
  T.~Liko,
  ``Topological deformation of isolated horizons,''
Phys.\ Rev.\ D {\bf 77}, 064004 (2008).
[arXiv:0705.1518 [gr-qc]].
}
\lref\HuangZUA{
  Y.~Huang and R.~X.~Miao,
  ``A note on the resolution of the entropy discrepancy,''
[arXiv:1504.02301 [hep-th]].
}
\lref\deHaroXN{
  S.~de Haro, S.~N.~Solodukhin and K.~Skenderis,
Commun.\ Math.\ Phys.\  {\bf 217}, 595 (2001).
[hep-th/0002230].
}

\lref\MyersTJ{
  R.~C.~Myers and A.~Sinha,
  ``Holographic c-theorems in arbitrary dimensions,''
JHEP {\bf 1101}, 125 (2011).
[arXiv:1011.5819 [hep-th]].
}

\lref\CoquereauxEX{
  R.~Coquereaux,
J.\ Algebra {\bf 398}, 258 (2014).
[arXiv:1209.6621 [math.QA]].
}

\lref\CoquereauxDW{
  R.~Coquereaux,
Rev.\ Union Mat.\ Argentina {\bf 51}, 17 (2010).
[arXiv:1003.2589 [math.QA]].
}

 \Title{} {\vbox{\centerline{Topological Entanglement Entropy,}
 \smallskip 
 \centerline{Ground State Degeneracy and Holography}
}}

\bigskip

\centerline{\it  Andrei Parnachev $^{1,2}$ and Napat Poovuttikul $^{2}$}
\bigskip
\smallskip
\centerline{${}^{1}$ School of Mathematics, Trinity College, Dublin 2, Ireland} 
\smallskip
\centerline{${}^{2}$Institute Lorentz for Theoretical Physics, Leiden University} 
\centerline{P.O. Box 9506, Leiden 2300RA, The Netherlands}
\smallskip

\vglue .3cm

\bigskip

\let\includefigures=\iftrue
\bigskip
\noindent
Topological entanglement entropy, a measure of the long-ranged entanglement,
is related to the degeneracy of the ground state on a higher genus surface.
The exact relation depends on the details of the topological theory.
We consider a class of holographic models where such relation 
might be similar to the one exhibited by 
Chern-Simons theory in a certain large $N$ limit.
Both the non-vanishing topological entanglement entropy and the ground state degeneracy 
in these holographic models 
are consequences of the topological Gauss-Bonnet term in the dual gravitational description.
A soft wall holographic model of confinement is used to generate  finite correlation length
but keep the disk topology of the entangling surface in the bulk, necessary for
nonvanishing topological entanglement entropy.

\bigskip

\Date{April 2015}

\newsec{Introduction and summary}

Entanglement entropy has recently emerged as an important quantity whose significance spans
various subjects ranging from quantum gravity to quantum Hall effect.
One of the properties of the latter is nonvanishing  topological entanglement entropy,
which can be defined to be a finite term $\gamma$  in a large radius expansion of the entanglement
entropy for a disk region in a 2+1 dimensional theory with finite correlation length  \refs{\LevinZZ,\KitaevDM}:
\eqn\eegentop{
S_{EE} = \alpha R - \gamma+ \ldots
}
More generally, topological entanglement entropy can be a good non-local order parameter for quantum liquids with 
long-range order in situations where more conventional local order parameters are not useful.

Many quantum Hall systems can be described by Chern-Simons theories, where topological 
entanglement entropy can be computed; it equals the $S_0^0$ component of the modular S-matrix of the theory \DongFT.
It can also be shown to represent a constant term in a partition function on a sphere or
a logarithm of the total quantum dimension.
Topological entanglement entropy is also related to the degeneracy of the ground states for theories
compactified on a spatial surface of genus $g$.
In particular, for abelian Chern-Simons theories, the relation is 
\eqn\abelianrel{
2 g \gamma = S_g\; ,
}
where $S_g$ is the logarithm of the number of the ground states. For a non-abelian Chern-Simons
theory, the relation between $S_g$ and $\gamma$ can be more
complicated.
As we review below, an interesting simplification occurs for the
$SU(N)_k$ Chern-Simons theory in the limit $N\gg k \gg 1$.
 In this case, the relation between $S_g$ and $\gamma$  is
\eqn\nonabelianrel{
2 (g-1) \gamma = S_g \; .
}

Recently, a seminal work by Ryu and Takayanagi \RyuBV\ has inspired work on 
entanglement entropy in the context of AdS/CFT correspondence. 
Although quite difficult to calculate by conventional field theoretic techniques, the entanglement entropy 
can be easily obtained in the dual holographic description by computing the minimal area of the bulk surface anchored on the entangling surface at the AdS boundary.
This raises a natural question whether one can construct
holographic theories with
  nonvanishing topological entanglement entropy $\gamma$.
Computing $\gamma$  involves finding an extremal surface in the bulk anchored to a large circle at the boundary of
 asymptotically AdS space.
In \PakmanUI\ this excersise was performed for the AdS-soliton geometry, which describes a confining 2+1 dimensional field
theory. 
In that model, at large radii, the dominant bulk hypersurface has the topology of a cylinder, and gives rise to the
vanishing topological entanglement entropy in accord with our expectations: such a QCD-like theory 
is not expected to have a ground state with nontrivial long-range order. For other work on topological entanglement entropy in the context of holography, see \refs{\FujitaKW,\McGoughGKA}.

In this paper we show that non-vanishing topological entanglement entropy arises naturally for certain field theories 
whose holographic dual four-dimensional descriptions include a Gauss-Bonnet term.
In four-dimensional gravity, the Gauss-Bonnet term is purely topological. Its contribution to the holographic entanglement 
entropy is also topological and can be computed by integrating the Euler density over the minimal surface in the bulk
 \refs{\FursaevIH\deBoerWK-\HungXB}. 
 It produces the Euler characteristic of that surface and hence, to get a non-vanishing $\gamma$, we need the minimal surface to have the disk topology  in the bulk, rather than the cylinder
topology.
This is precisely what did not happen in the example studied in \PakmanUI.
There are geometries where only a disk-like surface is a solution (and a cylinder-like is not) -- see e.g. \KulaxiziGY. 
The model considered in \KulaxiziGY\ does not quite work for us, since it does not have a gap in the excitation spectrum, but
it shows that we may have a chance by making the confinement ``softer".
Indeed, in this paper we show that certain soft-wall models of confinement do support entangling surfaces 
with the disk topology and, moreover, a constant term in the entanglement entropy due to the Einstein-Hilbert part of the bulk action is absent.
Hence, the addition of the Gauss-Bonnet term to these models ensures nonvanishing $\gamma$.

These four-dimensional holographic models (reviewed below) satisfy two important conditions: 
\item{(i)} The extremal surface anchored on a large circle at the AdS boundary has the topology of a disk.
\item{(ii)} The constant term in the large radius expansion of the area of this bulk surface vanishes.

\noindent The second condition comes from the expectation for $\gamma$ to come entirely from the Gauss-Bonnet term in the action, in the anticipation
of its relation to the  degeneracy of states on a genus $g$ surface.
This is because contribution of the Gauss-Bonnet term to $S_g$ is also topological, while a contribution 
from the Einstein-Hilber term to the entropy is necessarily extensive (equals the area of the horizon).
Indeed, we find that for the soft-wall models the relation between $\gamma$ and $S_g$ is  the same
as for the $SU(N)_k$ Chern-Simons theories  in the  $N\gg k \gg 1$ limit, \nonabelianrel.
To be more precise, this relation is only true for the part of the ground state entropy which comes from the
Gauss-Bonnet term.
As it turns out, in Einstein-Hilbert gravity it is very hard to find a holographic geometry, whose boundary is a genus $g>1$ surface,
with vanishing entropy at zero temperature.
This statement is true for pure AdS and remains true for models we consider.
This finite area of the horizon term spoils the relation  \nonabelianrel.
Such a term however is exponentially suppressed as the product of the
confining scale and the size of the genus $g$ surface becomes large. 

The rest of the paper is organised as follows.
 In Section 2 we review some facts about Chern-Simons theories.
 In particular we review relations \abelianrel\  and  \nonabelianrel.
 In Section 3 we review some basic facts about holographic four-dimensional Gauss-Bonnet gravity and point out 
 that a putative field theoretic dual of this gravitational model
 may possess non vanishing topological entanglement entropy  and ground state
 degeneracy, related by \nonabelianrel.
 In Section 4 we construct holographic models which have non-vanishing topological entanglement entropy -- the
 soft-wall models discussed above.
 In Section 5 we study the ground state entropy of these soft-wall models compactified
 on Riemann surfaces of genus $g$.
 We discuss our results in Section 6.
  The appendices A and B discuss the spectrum and the entanglement entropy of a slab region in the soft-wall model. We also discuss the topological entanglement entropy in  a certain holographic model of quantum Hall effect in appendix C.
\newsec{Quantum dimensions and ground state degeneracy in Chern-Simons theory}

In this section we briefly review the physical interpretation of the quantum dimension appearing in the Chern-Simons theory and its relation to the entanglement entropy. We finish the section by outlining how to obtain  the relation between topological entanglement entropy and ground state degeneracy on $\Sigma_g\times S^1$.

It is a well known fact that Chern-Simons theory is related to the corresponding Wess-Zumino-Witten (WZW) theory \WittenHF\foot{For an extensive review of WZW theory see e.g. \DiFrancescoNK.}.
Consider the theory on $T^2$. Under the modular S transformation, $\SS : \tau \to -1/\tau$, exchanging two non-contractable cycles, the WZW character is transformed as
\eqn\characterS{
\chi_a(-1/\tau) = \sum_b \SS^b_a\; \chi_b(\tau)\; .
}
The modular S-matrix, $\SS_a^b$ plays a crucial role in the corresponding Chern-Simons theory since the fusion rules \VerlindeSN\ (see also \NayakZZA\ for a review in condensed matter application) and partition functions on various Riemann surfaces \WittenHF\ can be written as combinations of these matrices.

We shall focus on the quantity called {\it quantum dimension}, $d_a = \SS_a^0/\SS_0^0$. This object has a special interpretation as a relative degeneracy of state in representation $a$ compared to the degeneracy of state in the identity representation, denoted as $a=0$ \DijkgraafTF. This can be shown more precisely by realising that the degeneracy of representation $a$ can be written as $\chi_a(q\to1)$, where $q=e^{2\pi i \tau}$. Thus, we have
\eqn\relativedim{
d_a=\lim_{q\to 1} \frac{\chi_a(q)}{\chi_0(q)} = \lim_{q\to 0}\frac{\sum_{b}\SS_a^b\chi_b(q)}{\sum_{b}\SS_0^b\chi_b(q)}= \frac{\SS_a^0}{\SS_0^0}
}
The topological entanglement entropy in \eegentop\  can be written as (se e.g. \DongFT)
\eqn\defqdim{
\gamma = -\log S_0^0=\log \DD \qquad ; \qquad \DD = \sqrt{\sum_a |d_a|^2} \; , 
}
where $\DD$ is called the {\it total quantum dimensions} \refs{\LevinZZ,\KitaevDM}. 
Recall also that $\SS_0^0$ is the partition function of the Chern-Simons theory on $S^3$ \WittenHF. 

We proceed by pointing out the relation between $\gamma$ and $S_g$ where $S_g$ is the entropy of the Chern-Simons theory on $\Sigma_g\times S^1$
\eqn\genusgdegen{
S_g =  \log \ZZ[\Sigma_g\times S^1]\; , 
}
where a simple relation \abelianrel\ can be found for $U(1)_k$ gauge group and \nonabelianrel\ for $SU(N)_k$ with $N\gg k \gg 1$
 \subsec{$U(1)_k$ abelian Chern-Simons theory}
 
For abelian Chern-Simons theories, the elements of the  S-matrix are just phases and, consequently, the total quantum dimension is simply
\eqn\totalqdimaCS{
\DD = \sqrt{\sum_{a=1}^{k}|d_a|^2} = \sqrt{k} \; , 
}
and, therefore, the topological entanglement entropy for this theory is
\eqn\eetopaCS{
\gamma = \log\,\DD = \frac{1}{2} \log (k)\; , 
}

 There are many ways to get the degeneracy of the $U(1)_k$ theory on a genus $g$ surface. 
 For the least mathematically involved approach, see \WenZG. The ground state degeneracy and the associated entropy are 
 \eqn\entpaCS{
 \ZZ[\Sigma_g\times S^1; U(1)_k] = k^g \qquad ; \qquad S_g = g \log k\; .
 }
 Thus, the topological entanglement entropy and ground state entropy are related by  \abelianrel.
 %
 %
 \subsec{$SU(N)_k$ Chern-Simons theory}
We now consider the non-abelian Chern-Simons theory with the $SU(N)_k$ gauge group where $N\gg k$.
The expression for the total quantum dimension can be found in e.g. \refs{\DongFT} but the calculation for large $N$ limit can be quite involved.
For an alternative method of computing total quantum dimensions,  see e.g. \refs{\CoquereauxDW,\CoquereauxEX}.
 A simple way to find a nice expression for the limit we are interested is to use the level-rank duality i.e. the partition function of $SU(N)_k$ and $SU(k)_N$ on $S^3$ are related by \CamperiDK 
\eqn\levelrankSph{
\frac{Z[S^3,SU(N)_k]}{Z[S^3,SU(k)_N]} = \sqrt{\frac{k}{N}}\; .
}
One can use the expression for $Z[S^3,SU(k)_N]$ for finite $k$ and $N\to \infty$, which is also presented in \CamperiDK,
\eqn\spherelargelevel{
\log \ZZ[S^3,SU(k)_N] \simeq -\frac{1}{2}(k^2-1)\log N + \OO(N^0) \; .
}
Using \levelrankSph\ and \spherelargelevel, the total quantum dimension $\DD$ in this limit is%
\eqn\qdimlargeN{
\DD = N^{+k^2/2} \qquad; \qquad \gamma =  \log \DD = \frac{k^2}{2}\log N \; .
}

As for the ground state degeneracy on $\Sigma_g\times S^1$, we use the same approach outlined above. The level-rank duality for this manifold is found in \IsidroEM\ to be 
\eqn\genusgSUNlevelrank{
\frac{\ZZ[\Sigma_g\times S^1, SU(N)_k]}{\ZZ[\Sigma_g\times S^1, SU(k)_N]} = \left( \frac{N}{k}\right)^g
\; . }
The expression for $\ZZ[\Sigma_g\times S^1, SU(k)_N]$ with $N\gg k$ can be found in \BanerjeeGH 
\eqn\SUNlargekgenusg{
\log \ZZ[\Sigma_g\times S^1,SU(k)_N] \simeq (g-1)(k^2-1)\log N + \OO(N^0)\; .
}
As a result, the entropy on $\Sigma_g\times S^1$, with the gauge group $SU(N)_k$, can be expressed as follows:
\eqn\entropygenusg{
S_g = \log \ZZ[\Sigma_g\times S^1,SU(N_k)] = g \log (N/k) + (g-1)(k^2-1)\log N + O(N^0)\; .
}
In the limit $N\gg k \gg 1$, we can then relate $\gamma$ and $S_g$ for $SU(N)_k$ using \qdimlargeN\ and \entropygenusg,
and the relation between them becomes \nonabelianrel.


\newsec{Gauss-Bonnet holography in 4 dimensions, entanglement entropy and ground state degeneracy on $\Sigma_g\times S^1$}
The Gauss-Bonnet theory is one of the simplest extensions of the Einstein gravity. It is described by the 
Einstein-Hilbert action with a 4-dimensional Euler density, the Gauss-Bonnet term, added. The action of this theory is
\eqn\defGBaction{
I = \frac{1}{16\pi G} \int d^4x \sqrt{-g} \left[ R + \frac{6}{L^2} +\frac{\lambda L^2}{2} E_4 \right]  \; ,
}
where
\eqn\defEulerden{
E_4 = R^2 - 4 R_{\mu\nu}R^{\mu\nu}+R_{\mu\nu\rho\sigma}R^{\mu\nu\rho\sigma}\; .
}
The Gauss-Bonnet term is non-dynamical, but affects physical quantities, such as the entropy and the entanglement entropy. The correction term is simply an additional constant proportional to $\lambda$. 
The Gauss-Bonnet gravity  is problematic in the following ways. It has been shown that for the positive Gauss-Bonnet coupling, $\lambda > 0$, one can merge black holes and violate the second law of thermodynamics \SarkarXP. 
We will mostly be interested in the situation with the negative 
coupling, $\lambda < 0$, where one can 
have black holes with negative entropy \refs{\JacobsonXS,\CveticBK,\ClunanTB,\LikoVI}. 
Also, graviton scattering in higher-dimensional Gauss-Bonnet theories exhibits violation of causality \CamanhoAPA. 
Nevertheless, none of these issues will appear in the present work.

In this section, we review computations of  the entanglement entropy and the black hole entropy in the presence of the Gauss-Bonnet coupling.
We show that  contributions from the Gauss-Bonnet term to the ground state entropy  and to the entanglement
entropy are related by \nonabelianrel. 
\subsec{Entanglement entropy and the entropy of topological black hole}

To calculate entanglement entropy for the field theory dual to 4-dimensional Gauss-Bonnet gravity, one has to find the minimum value of the following functional \refs{\FursaevIH,\deBoerWK,\HungXB}
\eqn\EEinGB{
S_{EE} = \frac{1}{4G}\left[{\int_M} d^2y \sqrt{\hat h} \left(1+\lambda L^2\hat{R} \right)+2\lambda L^2 \int _{\p M}dy \sqrt{\hat{h}_{\p}} \hat{K}\right] \; ,
 }
where $\hat{h}_{ij}$ is the induced metric on the minimum surface, $\hat{h}_\p$ is the induced metric of the boundary of the minimum surface, $\hat R$ is the Ricci scalar of the surface and $\hat K$ is the extrinsic curvature of the extremal surface's boundary. The same functional \EEinGB\ can also be obtained using the derivation of gravitational entropy in \refs{\BhattacharyyaGRA,\DongQOA,\CampsZUA,\BhattacharyyaYGA}, see also \refs{\MiaoNXA,\HuangZUA}.
In the conformal case, where the gravitational dual is the $AdS_4$ space, it is easy to compute the 
entanglement entropy of a disk of radius $R$.
 The minimum surface is described by $r(z)= \sqrt{R^2-z^2}$ and the entanglement entropy is
\eqn\EEdiscAdS{\eqalign{
S_{EE} &= \left[\frac{\pi L^2}{2G} (1-2\lambda) \int_{\epsilon}^R dz \left(\frac{R}{z^2}\right)\right] + 2\lambda\left(\frac{2\pi L^2}{4G}\right) \frac{R}{\epsilon}\cr
&= \frac{\pi L^2}{2G} \left(\frac{R}{\epsilon}-(1-2\lambda)\right)  \; .
}
}
Let us emphasise that, in general, there will be two kinds of constant terms in $S_{EE}$, similar to those in \EEdiscAdS. The first type of constant term, independent of the Gauss-Bonnet coupling, comes from the area of the minimum surface. However, this term is not topological as it receives corrections when the disc is deformed \refs{\HubenyRY, \AllaisATA}.
The constant term from the Gauss-Bonnet term, on the other hand, is proportional to the Euler characteristic and is topological by definition. As mentioned earlier, one of our main goals is to find a model where only the constant term of a second kind is nonzero.

The black hole entropy is no longer just the area of the horizon due to the presence of the Gauss-Bonnet term. 
The general formula for black hole entropy for the higher derivative gravity is the Wald entropy formula \IyerYS
\eqn\SWaldinGB{
S = \frac{1}{4G_N}\int_{\rm{horizon}}d^2y\sqrt{h}\frac{\partial \LL}{\partial R_{\mu\nu\rho\sigma}}\epsilon_{\mu\nu}\epsilon_{\rho\sigma} = \frac{1}{4G_N} \int_{\rm{horizon}} d^2y \sqrt{h} (1+\lambda L^2 \RR).
}
Here, $h_{ij}$ and $\RR$ are the induced metric and Ricci scalar on the black hole horizon. The Lagrangian density $\LL$ can be read off from the action \defGBaction. The binormal to the horizon $\epsilon_{\mu\nu}$ is defined as $\epsilon_{\mu\nu} = n^{(a)}_\mu n^{(b)}_\nu \epsilon_{ab}$, where $n_\mu^{(a)}$ are two unit normal vectors of the horizon and $\epsilon_{ab}$ is a usual Levi-Civita symbol. 

We are mostly interested in the constant term (the Euler characteristic) which is produced by integrating the Ricci scalar in \SWaldinGB\
over the hypersurface.
To double check our prescription and to ensure that no other constant terms are present, we can use the observation of
 \CasiniKV. 
 They showed that in a conformal theory, entanglement entropy of a disk equals the entropy of the hyperbolic black hole living in the dual AdS space.  To be precise, one can introduce the coordinate transformation 
\eqn\ztou{\eqalign{
L^2/z &=\rho\, \cosh (u) + \sqrt{\rho^2-L^2}\,\cosh(t/L),  \cr
L x^0/z &= \sqrt{\rho^2-L^2} \sinh(t/L),\cr
L x^1/z & = \rho \sinh(u)\cos\theta, \cr
L x^2/z & = \rho \sinh(u)\sin\theta \; .
}
}
This coordinate transformation maps the metric of the empty AdS 
\eqn\emptAdS{
ds^2 = \frac{L^2}{z^2} \left( dz^2 + \eta_{ij} dx^idx^j\right) \; ,
}
into the hyperbolic black hole with ${\bf R}\times H^2$ boundary, with a horizon\foot{
The appearance of the hyperbolic black hole here is similar to the way the Rindler space appears
once the Rindler coordinates are used to describe the wedge of the Minkowski spacetime.
See e.g. \MyersTJ\ for more detailed explanations.}
at $\rho =L$
\eqn\hypAdS{
ds^2 = \frac{d\rho^2}{\rho^2/L^2-1} - \left( \frac{\rho^2}{L^2}-1 \right)dt^2 + \rho^2 (d u^2 + \sinh^2 u\, d\theta ).
}
Using the Wald formula \SWaldinGB, the entropy of the hyperbolic black hole \hypAdS\ is 
\eqn\SWalddiscAdS{\eqalign{
S &= \frac{\pi \rho^2(1-2\lambda)}{2G} \int_{u=0}^{u_{\rm{max}}} du\; \sinh (u) \Big\vert_{\rho = L}\cr
&= \frac{\pi L^2}{2G}(1-2\lambda)\left( \frac{R}{\epsilon}-1\right) \; .
}
}
The value of $\cosh\, u_{\rm{max}} = R/\epsilon$ can be read off from the coordinate transformation \ztou\ at $\rho = L$.
 Now we can see that the physical, cutoff-independent, terms in \EEdiscAdS\ and \SWalddiscAdS\ are the same. This verifies the formula for the entanglement entropy in \EEinGB. 
%
\subsec{Gauss-Bonnet contributions to  entanglement entropy and to the black hole entropy with genus $g$ horizon}

The 2-dimensional Riemann surface of genus $g$ can be obtained by identifying the hyperbolic space $H^2$ by a finite subgroup of $H^2$ isometry \AminneborgIZ\ (see also \EmparanGF\ and reference therein). In this case the Gauss-Bonnet term in \SWaldinGB\ is proportional to the Euler characteristic, $\chi_g$, of the horizon due to the Gauss-Bonnet theorem\foot{
In the presence of boundaries this formula is modified; in particular, the  Euler characteristic of a disk is $\chi_{disk}=1$.}
\eqn\GBWaldgenusg{
S_{g}^{(1)} = \frac{\lambda L^2}{4G} (4\pi \chi_g) = \frac{2\pi \lambda L^2}{G} (1-g) \; ,
}
where  $S_g^{(1)}$ is the contribution to the entropy due to the Gauss-Bonnet term. Now, consider the Gauss-Bonnet term in the holographic entanglement entropy. 
The entangling surface can be found using the usual Ryu-Takayanagi prescription. 
In general, there are two types of surfaces that extremize the area functional, the ones with the cylinder topology with $r(z\to \infty) = const$ and 
those with the disc topology with $r(z=z_0)=0$, for some finite $z_0$. The Euler characteristic is zero for the cylinder and unity for the disc.
 Let us denote the Gauss-Bonnet contribution in the entanglement entropy by $S_{EE}^{(1)}$ .
  For the minimum surface with the disc topology, $S_{EE}^{(1)}$ can be written as
\eqn\GBEE{
S^{(1)}_{EE} = \frac{\lambda L^2}{4G} \int d^2 y \sqrt{\hat h}\hat R  = \frac{\pi \lambda L^2}{G},
}
where we used $\chi_{disk}=1$.
The expressions  \GBWaldgenusg\ and \GBEE\ are topological since they only depend on the topology of the horizon and minimum surface. 
The  relation between \GBWaldgenusg\ and \GBEE\ is the same as  \nonabelianrel.

\newsec{A soft-wall holographic model with nonvanishing topological entanglement entropy}

In this section, we consider  the geometry that is soft-wall  model \BatellZM\ in the IR and show that it
satisfies the criteria (i) and (ii) stated in the introduction. 
In the Appendix C, we also analyze the bottom-up model for fractional quantum Hall effect \LippertJMA\ in a certain range of parameters.
Without the Gauss-Bonnet term in the action, the topological entanglement entropy is zero. However, since (i) and (ii) are satisfied, these models may have nonzero topological term when the Gauss-Bonnet term is included.

The condition (i) will follow from the equation of motion, unaffected by Gauss-Bonnet term.
We will see that there is no minimum surface with cylinder topology for the large values of the radius $R$.
To ensure the condition (ii), one has to find the minimum surface and calculate the entanglement entropy to see that there is no $\OO(R^0)$ term in the area of the minimum surface when $R$ is large. 

We consider a class of geometries that 
are related to the AdS$_4$ by a warp-factor $a(z)$: 
\eqn\metrica{
ds^2 = \frac{L^2 a(z)}{z^2} \left( dz^2 - dt^2 +dr^2 +r^2d\theta^2\right)\; .
}
The equation of motion can be obtained by minimizing the area of the surface described by $r(z)$:
\eqn\Area{
A = 2\pi L^2 \int dz \frac{a(z)r(z)}{z^2}\sqrt{1+r'(z)^2}.
}
The equation describing the minimum surface is
\eqn\eoma{
\frac{a(z)}{z^2}\sqrt{1+r'(z)^2} = \frac{d}{dz} \Bigg( \frac{a(z)r(z) r'(z)}{z^2\sqrt{1+r'(z)^2}} \Bigg).
}
The metric is constructed to have a crossover scale between the UV and IR geometries
set by the mass scale, $\mu$. In the UV region, $\mu z \ll 1$, the warp factor is chosen to be 
\eqn\aUV{
a_{UV}(z \to 0) =1, \qquad a'_{UV}(z\to 0) = 0. 
}
The second condition in \aUV\ is chosen for technical convenience, so that the minimum surface of a disc radius $R$ near the boundary, $z/L\ll1$ is the same as in $AdS_4$:
\eqn\deepUVsurface{
r(z) = R- \frac{z^2}{2R} - \OO(1/R^2)
}

In the IR region, $\mu z \gg 1$, the warp factor is chosen to be the soft-wall warp factor of \BatellZM. 
\eqn\aIR{
a_{IR}(z) = e^{-(\mu z)^\nu}
}
The low energy spectrum of this theory can be found in existing works on soft-wall models. For $\nu =1$, the spectrum has a gap, set by the  energy scale $\mu$, and continuous spectrum above the gap \CacciapagliaNS. For $\nu >1$, the spectrum becomes gapped and discrete and for $\nu < 1$, the spectrum becomes gapless (see \refs{\BatellME,\FalkowskiFZ} and references therein). Note that, although the computations in the literature on soft-wall models are done in 5-dimensional gravity, the same conclusion can be reached in  4 dimensions, following the discussion in appendix A. The high energy spectrum of bound states in these models is given by
\eqn\glueballspectrum{
m^2_n = n^{2-2/\nu}
}
where $n$ is the excitation number of the bound states. For $\nu = 2$, one obtains a behaviour similar to the Regge trajectory in QCD \KarchPV. We will focus on the gapped system, $\nu \ge 1$, where the topological entanglement entropy is well defined.

As an explicit example, the warp factor that satisfies the above condtions is 
\eqn\explicitexample{
a(z) = 1/\cosh\left( \sqrt{(\mu z)^{2\nu} +1}-1 \right)
}
It is clear that the warp factor \explicitexample\ reduces to the IR form \aIR\ as $\mu z \gg 1$ 
(with the substitution $\sqrt{2} L \ra L$). 
In the UV regime, $\mu z \ll 1$, the warp factor above behaves as 
\eqn\explicitexampleUV{
a(z) \simeq 1- \frac{1}{8}(\mu z)^{4\nu}+\ldots
}
satisfying conditions in \aUV\ for all the range of $\nu$ where the spectrum is gapped, 
$1 \le \nu$ .
The profile of the surface with the warp factor in \explicitexample\ can be found numerically and the  relevant numerical results are shown in
   Figure 1.


We will now show that the surface with disc topology is the only possible solution when the surface probes the IR region,
as long as $\nu < 2$. The equation of motion describing the minimum surface in the IR can be written as 
\eqn\EoMdisc{
\frac{1}{z^2}e^{-(\mu z)^\nu}\sqrt{1+r'(z)^2} = \frac{d}{dz} \left( \frac{r(z)e^{-(\mu z)^\nu}}{z^2}\frac{r'(z)}{\sqrt{1+r'(z)^2}} \right)\; .
}
We will now employ the method outlined in the appendix C of \LiuEEA\ to rule out the solution with cylinder topology. For the cylinder topology solution to exist, there must be a solution to \EoMdisc\ with the asymptotic solution, $r(z\to \infty) = c_0$. Hence, the cylinder solution, at large $z$, must behave like
\eqn\rhozcylinder{
r(z) = c_0+c_1 z^m+\ldots \; ,
} 
with $m<0$. Note that $(\ldots)$ denotes the subleading term in $z\to \infty$ limit. Plugging in the ansatz \rhozcylinder\ into the l.h.s. of \EoMdisc\ and extracting the leading term, one finds that
\eqn\LHSEoM{
\frac{1}{z^2}e^{-(\mu z)^\nu}\sqrt{1+r'(z)^2}\; = \; z^{-2} e^{-(\mu z)^\nu} +\frac{m^2c_1^2}{2}z^{2m-4}e^{-(\mu z)^\nu}+\ldots \; .
}
Similarly to the r.h.s. of \EoMdisc, one finds that
\eqn\RHSEoM{\eqalign{
\frac{d}{dz} \left( \frac{r(z)r'(z)e^{-(\mu z)^\nu}}{z^2\sqrt{1+r'(z)^2}} \right) &=m c_0 c_1 z^{m-4} e^{-(\mu z)^\nu} \left(-\nu(\mu z)^\nu-3+m\right)
\cr
& + mc_1^2 z^{2m-4}e^{-(\mu z)^\nu}\left( -\mu (\mu z)^\nu-3+2m \right) + \ldots \; .
}}
We can see that, for the leading term on the left and right of \EoMdisc\ to match, we need the power of $z$ in these terms to be identical, namely
\eqn\condone{
z^{-2} = z^{ \nu+m-4} \quad\Rightarrow \quad \nu+m = 2 \; .
}
 However, for $\nu < 2$, we can see that \condone\ cannot be satisfied. 
Therefore, an extremal surface with the cylinder topology is not allowed for $\nu < 2$. 
Interestingly, this indicates that the phase transition for a disc region occurs precisely at $\nu = 2$, the point of linear confinement. 
On the other hand,  for the slab region (see  Appendix B), the phase transition occurs at $\nu =1$, where the  spectrum changes from  gapped to gapless.


We now proceed to show that the constant term in the large radius expansion of the entanglement entropy vanishes.
 This can be done by showing that 
the area of the minimum surface does not contain the $R^0$ term, where $R$ is the radius of the disc region at the boundary. 
We will consider the case of $\nu < 2$ where only the surface with the disc topology is a solution. 
The tip of the surface is located at $z=z_0$ and $\mu z_0 \gg 1$ at large $R$. 
The mass gap, $\mu$, is set to  unity for the rest of this section (equivalently, all dimensionful quantities
are measured in the units of $\mu$).
Following the method outlined in \LiuUNA, we split the minimum surface into three  parts:

\item{(I)} Deep UV region, $\epsilon < z < z^{(1)}_c$ , where $\epsilon$ is the UV cutoff: This  region contains part of the surface that attached to
the circle of radius $R$ at the boundary. The upper limit, $z^{(1)}_c$, is the crossover scale where $a(z)$ change the behaviour from $a_{UV}(z)$ to $a_{IR}(z)$. As $z \ll 1$, the minimum surface is described by \deepUVsurface\ as demanded by the construction of $a(z)$ in \aUV.

\item{(II)} Intermediate region, $z^{(1)}_c < z < z^{(2)}_c$ : In this region, the soft-wall warp factor, $a(z)$, becomes $a_{IR}(z)$ in \aIR. The upper limit, $z^{(2)}_c$, is chosen such that the area of the surface in this region is not exponentially suppressed by the warp factor $a_{IR}$ but will be suppressed when $z>z^{(2)}_c$. 
It is possible to find the profile of the minimal surface in the deep interior of this region i.e. when $z \gg 1$ but $z\ll z_0$. 
To do this, we introduce a new coordinate $u = z/z_0$ and $r(z) = (z_0)^n \rho(u)$. The power of $(z_0)^n$ is chosen such that $\rho(u)$ is of order $(z_0)^0$. The equation of motion \EoMdisc\ in this new parametrisation becomes
\eqn\eomregionII{
0 = -\frac{(z_0)^{2-2n}}{\rho(u)} - \nu u^{\nu-1} (z_0)^\nu \rho'(u) - \left( \frac{\rho'(u)}{\rho(u)} \right)^2 - \nu u^{\nu-1} (z_0)^{\nu-2+2n} (\rho'(u))^3 + \rho''(u).
}
Collecting leading terms in large $z_0$ expansion, one finds that the smallest value of $n$ that gives a nontrivial equation of motion is $n = 1- \nu /2$. For this value of $n$, the surface in this region is described by 
\eqn\solutionregionII{
\rho = \sqrt{\frac{2}{\nu(\nu-2)}\left( \nu(\nu-2) d_1 -u^{2-\nu}\right)},
}
where $d_1$ is an integration constant. Expanding the solution at small $u$ and writing it in the original parametrisation, one finds that
\eqn\expandsolutionregionII{
r(z) = \sqrt{2d_1} (z_0)^{1-\nu/2} - \frac{z^{2-\nu}}{\sqrt{2d_1}\nu(2-\nu)(z_0)^{1-\nu/2}} - \ldots
}

\item{(III)} Deep IR region, $z > z^{(2)}_c$ where $z$ can be of the same order as $z_0$ : as $z$ becomes very large, the area of the surface in this region is exponentially suppressed by the warp factor $a_{IR}(z)$.  

%
%
\ifig\loc{{\bf (LEFT) } Illustration of the region (I),(II) and (III).  {\bf (RIGHT) } Numerical value of radius $R$ versus the position of the tip $z_0$ with $a(z)^{-1} = \cosh(\sqrt{z^{2\nu}+1}-1)$. The gradient of this plot is $1-\nu/2$ for $\nu =1$(blue), $1.2$ (orange) and $1.4$ (green), respectively. }
{\epsfxsize2.2in\epsfbox{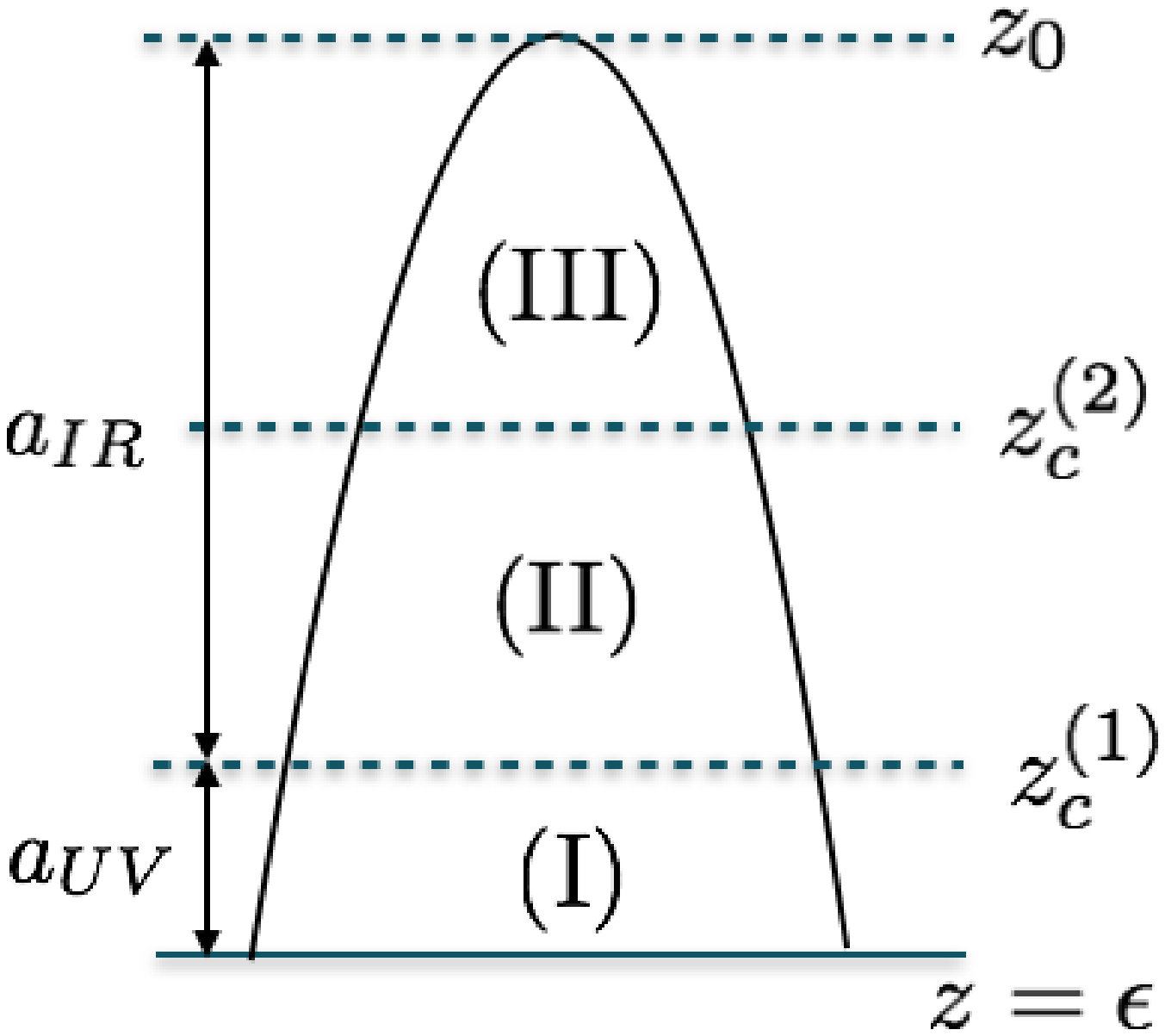}\epsfxsize3in\epsfbox{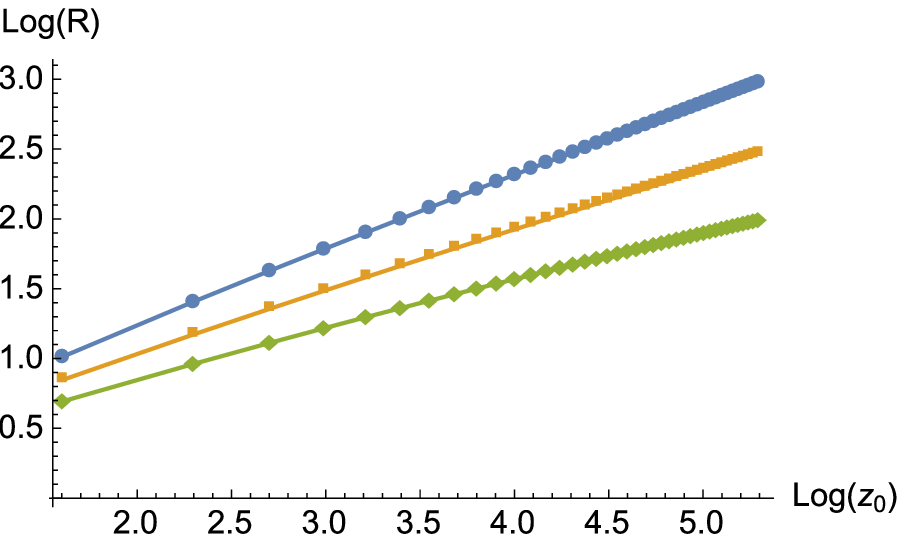}}
 From the solution $r(z)$ in region (I) and (II), one can see that the minimum surface can be described by the following large $R$ expansion,
\eqn\expandregionIandII{
r(z) = R - \frac{r_1(z)}{R} - \frac{r_2(z)}{R^2} + \OO(1/R^3),
}
where we identify $R\sim (z_0)^{1-\nu/2}$. The relation between $R$ and $z_0$ also agrees with the numerical results in figure 1. Here $\{ r_i(z)\}$ are functions interpolating between region (I) and (II). Note that this expansion breaks down  when $\nu$ approaches the critical value $\nu =2$. This is expected from the previous analysis since the minimal surface might change its topology at this critical value of $\nu$. 
Plugging the expansion \expandsolutionregionII, into the area of the minimum surface \Area, one finds  
\eqn\splitarea{\eqalign{
A &= A_{\rm(I)} + A_{\rm(II)} + A_{\rm (III)},
\cr
A_{\rm (II)} &= 2\pi L^2 \left[ R \int^{z^{(2)}_c}_{z^{(1)}_c}  dz \; \frac{a(z)}{z^2} + \frac{1}{R} \int^{z^{(2)}_c}_{z^{(1)}_c} dz \; \frac{a(z)}{z^2}\left( \frac{r_1'(z)^2}{2} - r_1(z)\right) + \OO(1/R^2)\right] , 
}}
where $A_{\rm (I)},A_{\rm (II)}, A_{\rm (III)}$ correspond to the area of regions (I),(II) and (III), respectively.  
In the region (I) the solution  is approximately  a cylinder of radius $R$ in the large $z_0$ limit, and yields a typical UV divergence $A\propto L^2R/\epsilon$ . The area of region (III) is exponentially suppressed by construction and can be neglected. 
To compute $A_{\rm (II)}$, we first note that the first cross over scale is of the order $1/\mu$ i.e. $z^{(1)}_c \sim 1$. Moreover, due to the fact that area from the region $z>z^{(2)}_c$ is negligible, the upper limit, $z^{(2)}_c$ can be lifted to infinity without drastically changing the integral $A_{(II)}$. Thus,  finite part of the area of the minimum surface can be written as
\eqn\Afinite{
A_{\rm finite} \approx 2\pi L^2 \left[ R \int^{\infty}_{1}  dz \; \frac{a(z)}{z^2} + \frac{1}{R} \int^{\infty}_{1} dz \; \frac{a(z)}{z^2}\left( \frac{r_1'(z)^2}{2} - r_1(z)\right) + \OO(1/R^2)\right] , 
}
and one can show that all integrals are  finite and independent of $R$. This indicates that there is no constant term, $R^0$, in the area of the minimum surface.

One can also check the validity of the above approximation scheme by computing the entanglement numerically using the warp factor \explicitexample. We found that the $R^0$ term in the area \Area\ has a value of order $10^{-4}$, for $1 \le \nu \le 2$ and  $\mu R\sim 20$.
This numerical value is negligibly small even though we are not at the limit $\mu R \gg 1$ and is expected to decrease even further as $\mu R$ increases.

Upon including the Gauss-Bonnet term, the entanglement entropy can be written as
\eqn\GBEEtop{
S_{EE} = \alpha_1 R+ \frac{\pi L^2\lambda}{G}  + \OO(1/R)\; ,
}
where $\alpha_1$ is some non-universal constant depending on the cutoff of the theory. Note that the boundary term in \EEinGB\ only gives a divergent term and therefore does not affect the R-independent piece. We can read off the topological entanglement entropy $\gamma$ 
by comparing \GBEEtop\ to \eegentop,
\eqn\gammaholo{
\gamma  = -\frac{\pi L^2 \lambda}{G} \; .
}

\newsec{Ground state degeneracy for the soft-wall model}

In this section, we make an attempt at computing
the ground state entropy for the model considered in the previous section
when  the horizon is the two-dimensional hyperbolic space.
As mentioned earlier, the surface of genus $g>1$ is obtained by identifying this hyperbolic space $H^2$ by a finite subgroup of the $H^2$ isometry. 
As in the empty AdS$_4$, the  area of the horizon is non-vanishing even at $T=0$. 
However, this area is suppressed 
when the product of the mass gap and the size, $L$, of $\Sigma_g$ is large\foot{It would be interesting to do the same calculation for the fractional quantum hall model of \LippertJMA. 
Unfortunately, unlike the flat horizon case, we find no hyperbolic black hole solution when the dilaton has a scaling form $\phi \sim r^N$ in the deep IR region for the allowed value of $\tilde\gamma$ and $s$}.

To find the black hole entropy with the horizon being $H^2$, one can proceed as the following. 
First, we replace the flat spatial metric by the $H^2$ metric. The black hole metric has the following form
\eqn\genbh{
ds^2 = \frac{L^2 a(z)}{z^2} \left( -f(z)dt^2 + \frac{dz^2}{f(z)} + d\Sigma_2^2\right)\; ,
}
where the spatial part of the boundary, $d\Sigma_2^2$, is the line element on $H^2$. 
\eqn\hyperbolictwometric{
d\Sigma_2^2 =L^2 \left( d\theta^2 + \sinh^2\theta d\phi^2\right) \; . 
}

As mentioned earlier, the black hole entropy contains a contribution from the Gauss-Bonnet term, even if the Gauss-Bonnet term is not dynamical. The entropy in this case can be computed using \SWaldinGB.
\eqn\SWald{
S = \frac{1}{4G}\left( \frac{L^2 a(z_H)}{z^2_H} \rm{vol}(\Sigma_g)\right) + \frac{\pi L^2}{G}\chi_H \; , 
}
where the first term on the most right hand side is the area of the horizon and $\chi_H$ is the Euler characteristic of the genus $g$ horizon i.e. $\chi_H = 2(1-g)$. 

In the following, we follow \HerzogRA, where a semi-quantitative method to 
construct the black hole solutions in the soft-wall models is proposed.
In this setup, the dilaton is assumed to be non-dynamical and does not affect the metric in the string frame 
\eqn\stringframe{
ds^2_{\rm{string}} = \frac{L^2}{z^2}\left( -f(z)dt^2 + \frac{dz^2}{f(z)}+d\Sigma_2^2 \right)\quad ; \quad ds^2 = e^{-2\varphi(z)}ds_{\rm{string}}^2 \; .
}
The dilaton $\varphi(z)$ is chosen such that it takes the form $\varphi(z) = -(\mu z)^\nu/2$, as in \refs{\KarchPV,\HerzogRA}.  
The emblackening factor is further assumed to be that of AdS-Schwarzschild
\eqn\emptyf{
f(z) = 1 - \frac{z^2}{L^2}+ M z^3 \; .
}
This solution is assumed to be a solution of a certain gravity model. To our knowledge, such model has not been found\foot{Given the metric of the form $ds^2 = e^{-2A(z)} (-dt^2+dx^2+dy^2+dz^2)$, one can try to use the potential reconstruction method \SkenderisMM\ to find a dilaton potential $V(\phi)$ when $A(z)= (\mu z)^\nu/2 + \log z$. The attempt to find $V(\phi)$ for black hole metric, $ds^2 = e^{-2A(z)}(-f(z)dt^2 + dx^2+dy^2 + dz^2/f(z))$, can be found in \LiHP. However, the dilaton potential for black hole phase is temperature dependent, so this route does not
work for us.
Another  way to construct the soft-wall black hole is considered in \KelleyVZ\ for a flat boundary and in the limit where the horizon is close to the boundary. In this method, the potential, $V(\phi,T)$, made out of dilaton, $\phi(z)$, and an additional scalar field, $T(z)$, is reconstructed from non-black hole geometry \BatellZM. To find a black hole solution, one has to solve for a nontrivial profile of $f(z), A(z),\phi(z)$ and $T(z)$. Unfortunately, we are unable to find an extremal hyperbolic black hole solution using this method.}.

The zero temperature solution can be found by tuning $M$ into $M_{ext} = (2/3^{3/2})L^{-3}$ and the horizon is located at $z _{H}= \sqrt{3}L$. As a result, the warp factor at the horizon can be written as 
\eqn\WarpNondyn{
\frac{L^2 a(z_H)}{z^2_H} = \frac{L^2}{z_H^2} e^{-\mu z_H} = \frac{e^{-\sqrt{3}\mu L}}{3} \; .
}
Substitute the expression in \WarpNondyn\ back into \SWald, one finds that the area of the horizon is exponentially suppressed when $\mu L \gg 1$.

Hence, in this limit, the entropy of the soft-wall theory on $\Sigma_g\times S^1$ contains only a Gauss-Bonnet term, which is topological
\eqn\SWaldgenusg{
S_g \approx  \frac{2\pi \lambda L^2}{G}(1-g) \; .
}
Relating the expression in \SWaldgenusg\ to the topological entanglement entropy is \gammaholo, we find the relation \nonabelianrel.
We emphasize that this result should be taken with a grain of salt, as a number of assumptions has been
made to arrive at \SWaldgenusg.
It would be great to find an honest way of constructing holographic gapped geometries with hyperbolic horizons.

\newsec{Discussion}

We show that it is possible to obtain nonvanishing topological entanglement entropy, $\gamma$, in holography.
The Gauss-Bonnet term plays a crucial role in our construction since $\gamma$ is 
proportional to the Gauss-Bonnet coupling.
The key property for the entangling surface to have a disk topology in the bulk is satisfied by
the soft-wall models we consider.
It is interesting that the soft and hard wall models of confinement are clearly distinct 
from the point of view of our work.
It would be interesting to identify a field-theoretic reason for this distinction.

Let us emphasize that the definition of $\gamma$ involves computing the entanglement entropy
of a disk in a theory on a plane. 
This is contrasted with a different measurement of the topological order: the degeneracy of the
ground state in the same theory compactified on a genus $g$ surface. 
We observe that the relation between the topological entanglement entropy and the contribution to the
ground state degeneracy from the Gauss-Bonnet term strongly resembles the same relation for
the Chern-Simons theory. It would be nice to understand this better.

To describe the holographic dual of the soft-wall model on a higher genus surface 
one needs to consider the bulk action that leads to the soft-wall metric
in the infrared and find the solution with the asymptotic boundary being $H^2 \times R$.
The analogous procedure in the conformal case (pure AdS) leads to the asymptotically 
AdS black hole with hyperbolic horizon, so the appearance of the horizon would not be surprising.
Unfortunately we did not succeed in constructing an honest soft-wall solution with a hyperbolic
horizon due to technical difficulties.
However, we present some arguments which indicate that  the horizon area (and therefore contribution 
of the Einstein-Hilbert term to the ground state degeneracy) is exponentially suppressed as the mass
gap becomes large.
It would be nice to make these arguments more precise.

To summarize, in this paper we showed how in certain holographic models with Einstein-Hilbert and
Gauss-Bonnet terms, the field theoretic degrees of freedom dual to the former are frozen in the infrared due to confining geometry,
while the latter presumably give rise to a topological theory.
To better understand the nature of this theory, more work is needed. In particular, considering holography
on spaces with boundary can provide new insights.

\bigskip
\bigskip
\noindent {\bf Acknowledgments:} We thank R. Davison, M. Freedman, M. Goykhman, N. Kaplis, Z. Komargodski, P. Kraus, M. Kulaxizi, S.-S. Lee, N. Mekareeya, G. Moore, S. Ross and K. Skenderis  for discussions. 
A.P. thanks Aspen Center for Physics and Simons Center for Geometry and Physics, Stony Brook University (Summer Simons Workshop 2014) for hospitality.
NP is supported by Leiden University and DPST scholarship from the Thai government and thanks Trinity College Dublin for hospitality. 
He also acknowledge the support during his visit at Trinity College Dublin, received from the European Science Foundation (ESF) for the activity entitled 'Holographic Methods for Strongly Coupled Systems'.
This work was also supported in part by the NWO Vidi grant and by the National Science Foundation  Grant No. PHYS-1066293 to the Aspen Center for Physics.

\appendix{A}{Spectrum of gauge invariant mode in soft-wall model}

In this section, we extract gauge invariant combinations of the metric fluctuations in the soft-wall model  and argue that the spectrum of the metric fluctuations in the 4-dimensional model is gapped as in the 5-dimensional one. 

The spectrum of the metric fluctuation of the 5-dimensional soft-wall model with general $\nu$ was studied in \refs{\BatellME}. Nevertheless, the metric fluctuations in 5 and 4 dimensions are slightly different. The 5-dimensional metric fluctuations, $h_{\mu\nu}(x^0,x^1)$, can be categorised into three different channels by the remaining $O(2)$ symmetry of the boundary field theory (see e.g. \KovtunEV). In this case, it is enough to consider the scalar channel, consisting of only $h_{x^2x^3}$, and solve scalar field equation of motion, where $x^i$ denote the directions along the boundary. On the other hand, in 4-dimensional gravity, the fluctuation is categorised by the parity, $y\to -y$ (see e.g. \DavisonBXA). As a result, the odd parity channel contains $(h_{yt},h_{xy},h_{zy})$. The rest of the metric and scalar fields fluctuations are in the even parity channel. The fluctuations are coupled to components in their respective channel and, a priori, it is not clear that the analysis in 5-dimensional theory is still valid.

We follow the approach similar to those in \DavisonBXA\ by finding a gauge invariant combination of the fluctuation in odd parity channel. The equations of motion of these fluctuations are 
\eqn\fluctuation{\eqalign{
(\omega^2-k^2)h_{zy}(z) - ik h'_{xy}(z)+\omega h'_{yt}(z) &=0\,,\cr
\frac{z}{L}a(z)^{-1/2}\frac{d}{dz}\left[\frac{L}{z} a(z)^{1/2}(h'_{xy}(z)-ik h_{zy}(z)) \right]+\omega\left( \omega h_{xy}(z)+kh_{yt}(z)\right) &=0,\cr
\frac{z}{L}a(z)^{-1/2}\frac{d}{dz}\left[\frac{L}{z} a(z)^{1/2}\left(h'_{yt}(z)+i\omega h_{zy}(z)\right)\right]-k\left(kh_{yt}(z)+\omega h_{xy}(z)\right)&=0.
}}
where $a(z)= e^{-(\mu z)^\nu}$ is the warp factor of the soft-wall metric \metrica\ in the IR. These three equations of motion are not independent. We can rearrange the last two equations into the equation of motion of the gauge invariant combination by eliminating $h_{zy}(z)$. The resulting equation is
\eqn\gaugeinvfluc{
\frac{z}{L}a(z)^{-1/2}\frac{d}{dz}\left[\frac{L}{z}a(z)^{1/2}\varphi'(z) \right] + (\omega^2-k^2)\varphi(z) = 0\, , \qquad \varphi(z) = h_{xy}(z)+ \frac{k}{\omega}h_{yt}(z) \; .
}
The equation of motion above is identical those of the KK modes in 5-dimensional soft-wall \FalkowskiFZ.
(see also \CacciapagliaNS\ and \BatellME\ for discussions in $\nu=1$ and $\nu=2$ case). Thus, we conclude that the spectrum of the metric fluctuations in the 4-dimensional soft-wall is also gapped as in the 5-dimensional one.

\appendix{B}{Entanglement entropy for slab geometry}
In this section, we study the entanglement entropy for the slab region in the field theory dual to the model described by the metric \metrica. 
The case of soft-wall model with $\nu =2$ was briefly mentioned in \KlebanovWS\ and further studied in \KolNQA. See also \refs{\AmiGSA}, for multiple slab regions in different gap phases. We found that the phase transition of the entanglement entropy occurs at $\nu=1$, precisely the value when the spectrum changes from being gapped to gapless \refs{\BatellME,\FalkowskiFZ}. 

Let us introduce the notation here. The slab region is the region between $x = \pm \ell/2$ and has an infinite length along $y-$direction. The induced metric on the minimum surface is written as  
\eqn\inducedslab{
ds^2_{\rm{ind}} = \frac{L^2a(z)}{z^2} \left( (1+z'(x)^2)dx^2 + dy^2  \right) 
}
To be consistent with the main text, the numerical result in this appendix is done with the warp factor $a(z) = 1/ \cosh (\sqrt{z^{2\nu}+1}-1) $, where we set the energy gap scale $\mu =1 $ for simplicity.
We will first look at the profile of the minimum surface, determined by the area functional. 
The area of the surface in this section can be written as
\eqn\actionslabnu{
A = L^2 \int_{-\infty}^{\infty} dy\int_{-\ell/2}^{\ell/2} dx \sqrt{H(z)}\sqrt{1+z'(x)^2}\quad ; \quad H(z) = a(z)^2/z^4 \; ,
}

As pointed out in previous studies, there are two possible configurations that satisfy the equation of motion derived from \actionslabnu. The first solution is two infinitely long parallel planes described by $x(z) = \pm \ell/2$, referred to as disconnected surfaces. The other solution is the connected surface where $z'(x)=0$ at some value of $z_0$. To find the profile of $z(x)$ describing the latter solution, we notice that area \actionslabnu\ does not explicitly depend of $x$ and one can use the ``conservation of energy'' to obtain the following first order equation 
\eqn\EoMslapnu{
\frac{H(z)}{1+z'(x)^2} = H(z_0).
}
Inverting this equation of motion, one can obtain the relation between $z_0$ and the width $\ell$ of the slab as 
\eqn\ellzm{
\ell = \int_{-\ell/2}^{\ell/2}dx = \int_{0}^{z_0} dz \frac{\sqrt{H(z_0)}}{\sqrt{H(z)-H(z_0)}}  ,
}
The profile of $\ell$ in \ellzm\ as a function of $z_0$ is shown in figure 2. We found that for $\nu > 1$, the width $\ell(z_0)$ has a maximum value, $\ell = \ell_{\rm max}$. This indicates that there can only be the disconnected surfaces when $\ell > \ell_{\rm max}$. The maximum value $\ell$ is not found for $\nu < 1$.   

\ifig\loc{The width of the extremal surface $\ell(z_0)$ as a function of $1/z_0$. The width $\ell(z_0)$ has a maximum value at large $z_0$ for $n>1$. The maxima of $\ell(z)$ is not found for $\nu < 1$. This plot is obtained numerically from \ellzm}
{\epsfxsize3.5in\epsfbox{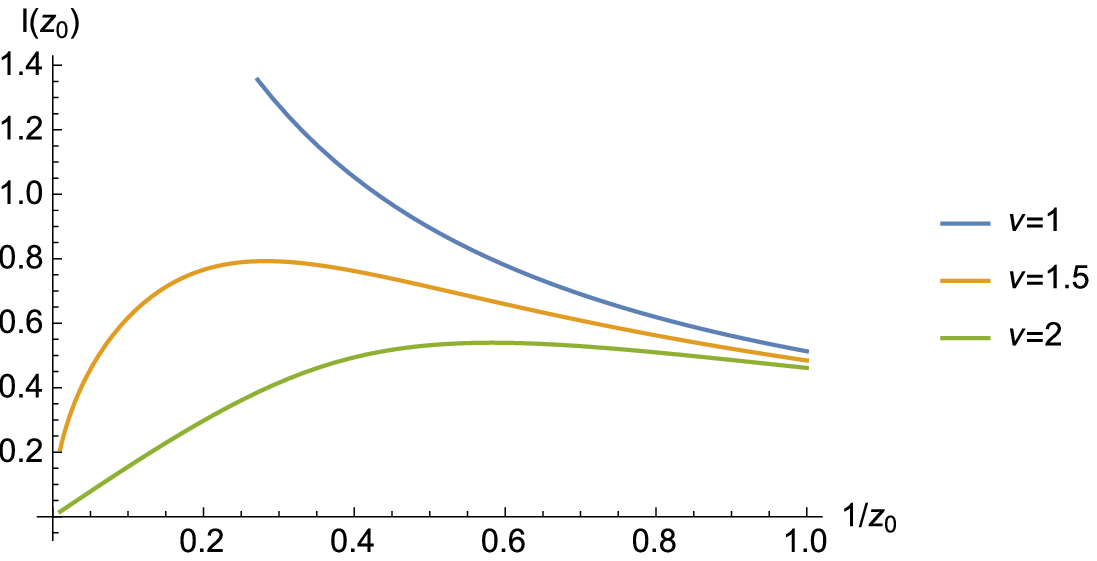}}
To find the entanglement entropy, one needs to calculate the area \actionslabnu\ for both solutions and finds out which one is smaller. It turns out that the Gauss-Bonnet contribution for both solutions are zero and one can obtain the entanglement entropy by simply computing the area.
 The area of both surfaces can be found by numerically evaluate the following integrals
\eqn\Aconslab{\eqalign{
s^{\rm{con}} &= L^2\int_{\epsilon}^{z_0} dz \frac{H(z)}{\sqrt{H(z)- H(z_0)}},\qquad
s^{\rm{discon}} = L^2\int_{\epsilon}^\infty dz \sqrt{H(z)} \; .
}}
Here $\epsilon$ denotes the short distance cutoff while $s^{\rm con}$ and $s^{\rm discon}$ represent the areas divided by length along $y-$direction of connected and disconnected surface respectively. 
The difference between the areas of connected and disconnected surfaces, $\Delta s= s^{\rm{con}}-s^{\rm{discon}}$, for $\nu = 1$ and $1.5$ are shown in figure 3. We can see that, for $\nu = 1$, $\Delta s$ approaches zero from below as we increase the width of the slab. Hence, for a slab region with a finite width, the connected surface remains a preferable solution. This is also true for the theory with $\nu < 1$. For $\nu > 1$, the minimum surface at small $z_0$ is the connected surface but undergoes the phase transition into disconnected surfaces at large $z_0$, as depicted in figure 3 for $\nu =1.5$. 

To sum up, the entanglement entropy for a slab region with a large width $\mu \ell \gg 1$ is governed by disconnected surfaces for $\nu >1$ when the theory is gapped and by connected surface for $\nu < 1$ when the theory is gapless. 
As in \KulaxiziGY, the critical point $\nu=1$,where the phase transition occurs  does not affect the entanglement entropy
of a disk.
This seems to indicate that the simple model of \KlebanovWS\ actually needs more work.
\ifig\loc{The difference of the areas, $\Delta s$, of the surfaces described by \Aconslab\ in the unit of $L^2$as a function of $1/z_0$. We can se that for $\nu = 1.5$, there is a region where $\Delta s > 0$ indicated that the entanglement entropy is governed by disconnected surfaces when $z_0$ is large. For $\nu=1$, the connected surface gives the smaller area for any $z_0$. }
{\epsfxsize3.5in\epsfbox{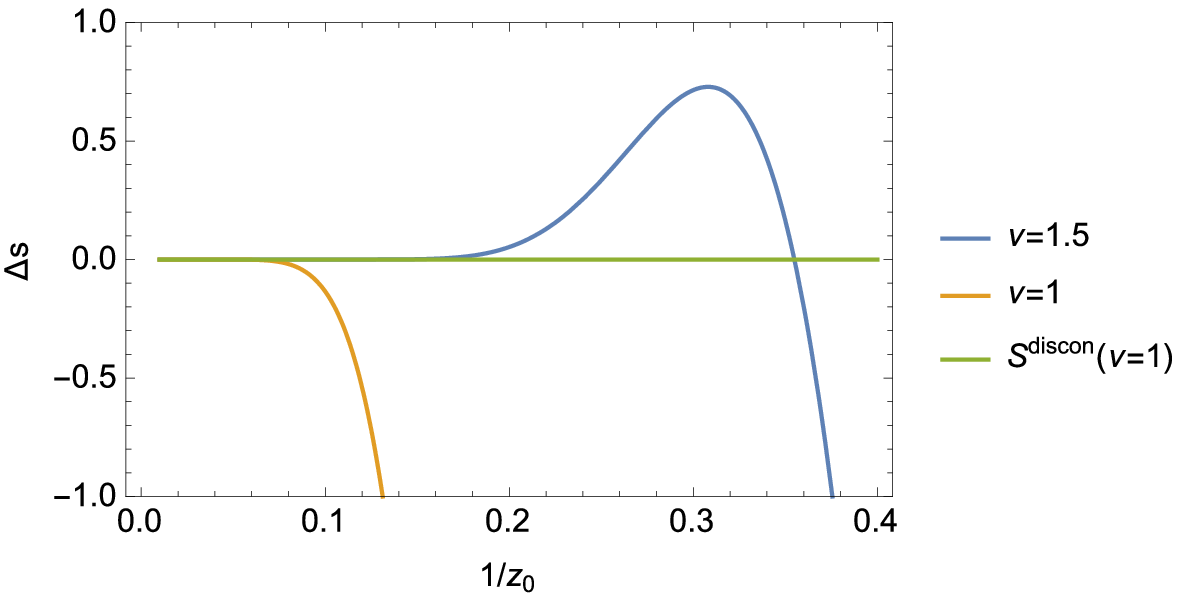}}


\appendix{C}{Bottom-up model of fractional quantum Hall system}

In this section, we apply our procedure to a bottom-up model of quantum Hall system \LippertJMA. We show that this model has non-zero topological entanglement entropy at certain values of free parameters $(\tilde \gamma, s)$ in the action of \LippertJMA\ when the Gauss-Bonnet term is present\foot{In \LippertJMA\ the two free parameters are denoted as $\gamma$ and $s$. We denoted their $\gamma$ by $\tilde \gamma$ to a avoid confusion with the topological entanglement entropy}. 

Let us briefly review the setup for this model. The action we consider here is the extension of Einstein-Maxwell-dilaton-axion theory, with four following components in the action
\eqn\actionQHE{
I = I_g + I_F +I_V + I_{GB} \; ,
}
where $I_{GB}$ is the usual Gauss-Bonnet term defined in \defGBaction. Note again, that 
$I_{GB}$ is does not affect the analysis in \LippertJMA\ and can be added to produce the topological entanglement entropy. The pieces $I_g$,  $I_F$ and $I_V$ can be written as
\eqn\actionSgSF{\eqalign{
I_g & = \int d^4x \sqrt{-g} \left[ R - \frac{1}{2} (\partial \phi)^2 - \frac{1}{2}\frac{e^{-2\tilde\gamma \phi}}{\tilde\gamma^2} (\partial \tau_1)^2 \right] ,\cr
I_F & = -\frac{1}{4}\int d^4x \left[\sqrt{-g} e^{\tilde\gamma \phi} F^2 +\frac{\tau_1}{2}\tilde\epsilon^{\mu\nu\rho\sigma}F_{\mu\nu}F_{\rho\sigma} \right], \cr
I_V & = \int d^4x \sqrt{-g} \,V(\phi,\tau_1)\; ,
}
}
where $V(\phi,\tau_1)$ is the $\rm{SL}(2,{\bf Z})$ invariant potential of the axion $\tau_1$ and dilaton $\phi$. The solution we are interested in has both electric and magnetic fields however the electric field is completely screened out in the deep IR. In this region, the potential $V(\phi,\tau_1)$ is approximately $V = -2\Lambda e^{-\tilde\gamma s\phi }$ and the extremal solution can be written as hyperscaling violating Lifshitz geometry
\eqn\IRQHEmetric{
ds^2 = C(\tilde\gamma,s,h) \rho^\theta \left[ -\frac{dt^2}{\rho^{2\tilde z}}+\frac{d\rho^2 + dx^2+dy^2}{\rho^2}\right] \; . 
}
The overall constant, $C(\tilde\gamma,s,h)$, depends on the free parameters in the action and the total magnetic field $h$. The dynamical exponent, $\tilde{z}$, and hyperscaling violating exponent, $\theta$, can be written in terms of $\tilde \gamma$ and $s$ as 
\eqn\zandtheta{
\tilde z = \frac{\tilde\gamma^2(1+s)(1-3s)+4}{\tilde\gamma^2(1+s)(1-s)} \quad ; \quad \theta = \frac{4s}{s-1}
}
where $\rho \to \infty$ corresponds to the AdS boundary. We can define a new coordinate $\rho = \left( \frac{2}{\theta-2}\right)^{2/\theta} z^{2/(2-\theta)}$ so that the boundary is at $z\to 0$ when $\theta >2$. The induced metric on the entangling surface in this new coordinate is 
\eqn\inducedQHE{
ds^2_{\rm{ind}} = \frac{1}{z^2} \left( \left(1+ \frac{z'(r)^2}{z^n} \right) dr^2+ r^2d\Theta^2  \right) , \qquad n = \frac{2\theta}{\theta-2}\; ,
}

The allowed value of $\theta$ (or equivalently $\tilde \gamma$ and $s$) can be found by imposing consistency conditions on the potential $V(\phi,\tau_1)$ and by demanding that the theory is gapped and has no naked singularity at finite temperature. The results in \LippertJMA\ show that the allowed values of $\gamma$ and $s$ are the small region around the line $s = 1\pm 1.44 (\tilde\gamma\pm 1)$ between $\tilde\gamma \in \pm [0.75,1]$. Since the width in $\tilde\gamma$ direction is small, we approximate the allowed region of $(\tilde\gamma,s)$ to be a straight line depicted in figure 4.
\ifig\loc{({\bf LEFT}) The allowed value of $s$ and $\tilde\gamma$ captured by the relation $s= 1\pm 1.44(\tilde\gamma\pm 1)$ where $\tilde\gamma \in \pm [0.75,1]$ ({\bf RIGHT}) The allowed value of $n = 2\theta/(\theta-2)$ from value for allowed value of $\tilde\gamma$ and $s$. Noted that $n=2$ is the minimum value when $s=1$ and $\tilde\gamma = \pm 1$}
{\epsfxsize2.5in\epsfbox{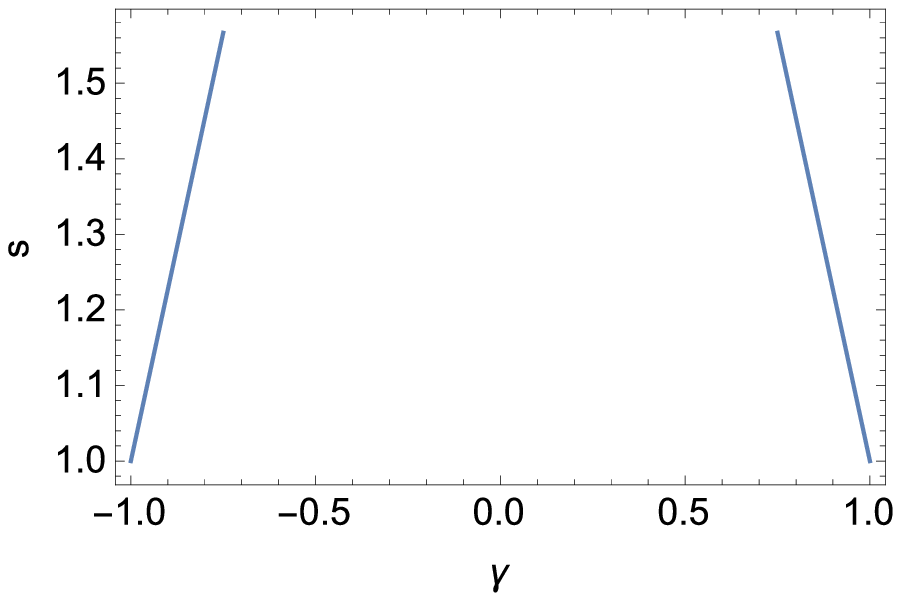}\epsfxsize2.5in\epsfbox{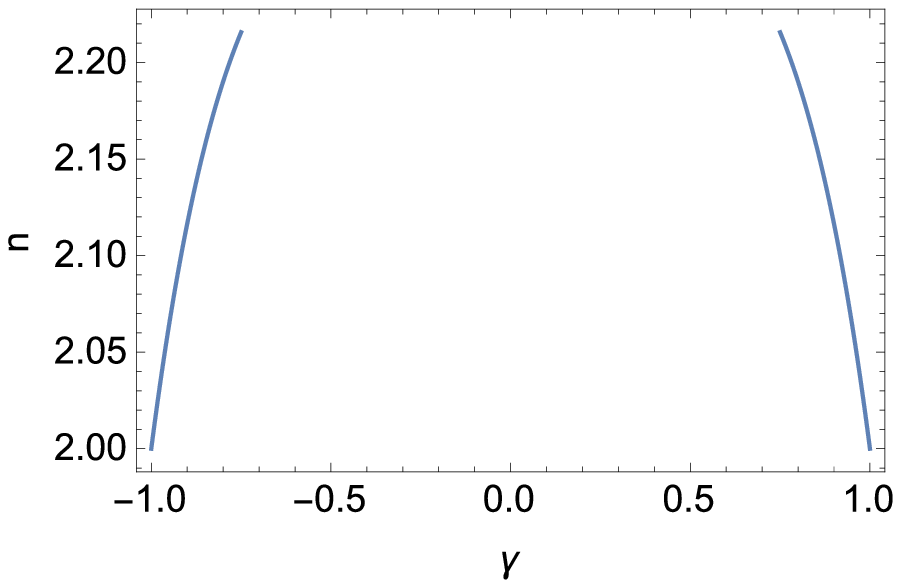}}

The entanglement entropy calculated from the induced metric of the form \inducedQHE\ has been explored in \LiuEEA. One of their results is that there is only a solution with a disc topology when $n \le 2$. Thus, from condition (i), only theory with $\tilde\gamma=\pm 1$ and $s=1$ can have non-vanishing topological entanglement entropy. 

Moreover, when $n=2$, the approximate solution of the entangling surface can be found. 
Let us parametrise the extremal surface by $z=z(r)$, the equation of motion can be written as
\eqn\eomFQHE{
\frac{d}{dr}\left( \frac{r z'(r)}{z(r)^4\sqrt{1+ (z'(r)/z(r))}}\right) = - \frac{2r (z(r)^2 + 3 z'(r)^2)}{z(r)^5 \sqrt{1+ (z'(r)/z(r))^2}}
}
We are interested in the limit where the crossover from the deep IR geometry to the full metric happens at $z=z_c(R_c)$ and that $R_c\gg 1$, in the unit of AdS radius. In the limit where $r$ approaches zero, one finds that the ansatz $z(r) \approx Z e^{- A r^2}$ solves the equation of motion \eomFQHE\ at the leading order. After imposing the matching condition, $z(R_c) = 1$, to fix the constants $Z$ and $A$, we find that the surface in the region $z> z_c$ is described by
\eqn\approxsolQHE{
z(r) = \exp\left( \frac{R^2-r^2}{2}\right)
}
Now, we need to see whether there is a nonzero constant term from the area of the extremal surface or not. We assume that the entangling surface extend very deep in the IR so that the finite part of the area is determined by the IR part, similar to examples in section 4 and in \LiuUNA. Also, in the large $R$ limit, the crossover radius $R_c$ is approximately equal to $R$, where $R$ is the radius of the surface at the AdS boundary. In the large $R\approx R_c$ limit, we have 
\eqn\areaQHE{\eqalign{
S_{RT} &= \frac{\pi}{2G}\int dr \frac{r}{z(r)^2} \sqrt{1+\frac{z'(r)^2}{z(r)^2}}  \cr
&\approx \frac{\pi}{2G}\int_0^{R_c\approx R} dr r^2 e^{\frac{r^2-R^2}{2}} \approx  \frac{\pi}{4G}\left(R -\frac{1}{R}\right)
}
}
This indicates that there is no constant term from this geometry. Hence, in this approximation, this model passes the criteria (i) and (ii) mentioned in the introduction. In principle, one should also extract the constant term in the UV completed metric, not just the IR part. However, the UV completed metric in \LippertJMA\ has to be obtained numerically which is beyond the scope of this work.

 \footatend\vfill\supereject\immediate\closeout\rfile\writestoppt
\baselineskip=14pt\centerline{{\bf References}}\bigskip{\frenchspacing%
\parindent=20pt\escapechar=` \input refs.tmp\vfill\eject}\nonfrenchspacing
\end